\documentclass[12pt,fleqn,leqno]{article}
\usepackage{cauchy}
\usepackage{amsmath,amssymb,amsbsy}
\usepackage{subfigure}
\usepackage{graphicx}
\usepackage{multirow}
\usepackage{lscape}
\usepackage{slashbox}

\renewcommand{\baselinestretch}{1.2}

\makeatletter
\@addtoreset{equation}{section}
\renewcommand\thetable{\@arabic\c@table}
\renewcommand\thefigure{\@arabic\c@figure}
\long\def\@makecaption#1#2{%
  \vskip\abovecaptionskip
  \begin{center}%
  \sbox\@tempboxa{#1: #2}%
  \ifdim \wd\@tempboxa >\hsize
    #1: #2\par
  \else
    \global \@minipagefalse
    \hb@xt@\hsize{\hfil\box\@tempboxa\hfil}%
  \fi
  \end{center}%
  \vskip\belowcaptionskip}
\def\N{{\rm I\kern-.15em N}}
\def\R{{\rm I\kern-.2em R}}
\def\Z{{\rm Z\kern-.26em Z}}
\makeatother

\newtheorem{thm}{Theorem}[section]
\newtheorem{lem}[thm]{Lemma}

\newcommand{\be}{\begin{eqnarray}}
\newcommand{\ee}{\end{eqnarray}}
\newcommand{\bq}{\begin{eqnarray*}}
\newcommand{\eq}{\end{eqnarray*}}

\begin{document}
\begin{center}
{ \Large Data Transformations and Goodness--of--Fit Tests for Type--II Right Censored Samples       \\}
\vspace*{0.5cm}
{\bf Christian Goldmann$^a$, Bernhard Klar$^b$\footnote{Corresponding author. {\it Tel.} +4972160842047 \\
 {\it Email address:} bernhard.klar@kit.edu}, Simos G. Meintanis$^{c,d}$} \\

\vspace*{0.5cm}

{\it $^a$Fraunhofer Institut f\"ur Techno- und Wirtschaftsmathematik ITWM, \\ Frauhofer Platz 1, 67663 Kaiserslautern, Germany

$^b$Department of Mathematics, Karlsruhe Institute of Technology (KIT), \\ Kaiserstra{\ss}e~89, 76133 Karlsruhe, Germany

$^c$Department of Economics, National and Kapodistrian University of Athens, \\ 8 Pesmazoglou Street, 105 59 Athens, Greece

$^d$Unit for Business Mathematics and Informatics, North--West University, Potchefstroom, South Africa}

\end{center}


\vspace*{1cm} \noindent
{\small {\bf Abstract.} We suggest several goodness--of--fit  methods which are appropriate with Type--II right censored data. Our strategy is to transform the original observations from a censored sample into an approximately i.i.d. sample of normal variates and then perform a standard goodness--of--fit test for normality on the transformed observations. A simulation study with several well known parametric distributions under testing reveals the sampling properties of the methods. We also provide theoretical analysis of the proposed method.

\vspace*{0.5cm} \noindent

\newpage


\section{Introduction}\label{sec_1}

 In a sample of size $n$ from the distribution, assume that only the first $r\leq n$ order statistics  $X_{1:n}<X_{2:n}<\ldots<X_{r:n}$, are observed. This censoring scheme is referred to as Type--II censoring. Let $X$ be the underlying random variable and denote by $F(x)$ the distribution function (DF) of $X$. We are interested in the goodness--of--fit (GOF) null hypothesis \be H_0: F \equiv {\cal {F}}_{\vartheta}, {\rm{for \ some}} \  \vartheta \in \Theta,\ee with $\Theta \subseteq \R^p, \ p\geq 1$, where ${\cal {F}}_{\vartheta}$ denotes a specific family of distributions indexed by a parameter $\vartheta$. Typically the null hypothesis in (1.1) is tested by modifications of the standard GOF tests. Early works include the Cram\'er--von Mises statistic by Pettitt (1976, 1977), and the Kolmogorov--Smirnov test by Barr and Davidson (1973) and Dufour and Maag (1978), all with the assumption that the parameter $\vartheta$ is known, and the chi--squared tests with estimated parameter in Mihalko and Moore (1980). Other less conventional approaches are the regression tests for exponentiality of Brain and Shapiro (1983), the smooth tests of fit of Bargal and Thomas (1983) and tests based on normalized spacings suggested by D'Agostino ansd Massaro (1991). A standard reference for GOF tests, including tests with censored data, is D'Agostino and Stephens (1986), while Thode (2002, Chapter 8) contains a nice overview of  various methods of testing normality with type--I and type--II censored data. For recent approaches to GOF tests with censored data the reader is referred to Gran\'e (2012), Castro--Kuriss (2011), Castro--Kuriss et al. (2010), Glen and Foote (2009), and Pe$\tilde{\rm{n}}$a (1995), among others.

As already noted the standard approach for the testing  problem in (1.1) has been to consider test statistics for the case of no censoring $r=n$, and modify them accordingly in order to make them applicable for the case $r<n$.  At the same time however, there exist methods which, given the $r$ order statistics in a random sample of size $n$ from the uniform (0,1) distribution, and based on specific transformations of the data, yield order statistics in a random sample of size $r$ from the same distribution. Then of course any GOF statistic for uniformity may be applied as if we had a full sample of size $r$ to begin with.  Such `transformations--to--uniformity' appear in Michael and Schucany (1979), O'Reilly and Stephens (1988), and more recently in Lin et al. (2008), and Fischer and Kamps (2011). Clearly testing uniformity is not a restriction since these statistics naturally extend to the current setting of arbitrary null hypothesis $H_0$ by use of the probability integral transform.

There is also a line of research which can be combined effectively with the aforementioned transformations--to--uniformity. In particular, and since under $H_0$ the parameter $\vartheta$ needs to be estimated from the data, we essentially have a quasi--probability integral transform, with extra variability introduced during the estimation step. Consequently, the corresponding GOF statistics will depend on the unknown value of the parameter and the method of estimation used in estimating this parameter. In this connection, and in order to make the GOF statistics independent of these choices, Chen and Balakrishnan (1995) proposed a novel transformation--to--normality for this problem. As a result, and by combining the Chen--Balakrishnan  transformation--to--normality with any of the aforementioned  transformations--to--uniformity, we can conveniently reduce any given testing problem with Type--II censoring to a GOF test for normality with complete samples, which of course is a well studied problem with many solutions. We note that the empirical process underlying the Chen--Balakrishnan transformation was first analysed in the PhD thesis by Chen (1991). Here we provide further theoretical as well as empirical results justifying the general validity of the this transformation.

In this paper we apply these transformations--to--uniformity in conjunction with the Chen--Balakrishnan transformation to several GOF tests.  The rest of the paper unfolds as follows. In Sections 2 and 3 we present the transformations and indicate how to implement them in the corresponding GOF statistics. Section 4 deals with the issue of estimating the parameter $\vartheta$ under Type--II censoring. In Section 5 a Monte Carlo study is drawn in which several combinations of tests statistics and transformations are studied in their sampling properties. Finally Section 7 contains the conclusions of this study. A theoretical analysis of the basic process involved in the Chen--Balakrishnan transformation, assisted by simulations, is provided in the Appendix.

\section{Transformations}\label{sec_2}

Denote by $U(0,1)$ the uniform distribution on (0,1) and suppose that $U_{1:n}<U_{2:n}<\ldots<U_{r:n}$, are the first $r$ order statistics in a random sample of size $n$ from $U(0,1)$ distribution. Further, put $U_{0:n} \equiv 0$.
Let ${\bf{U}}:=(U_{1:n},U_{2:n}\ldots U_{r:n})^{\rm{T}}$ and denote by ${\bf{u}}:=(u_{1:r},u_{2:r}\ldots u_{r:r})^{\rm{T}}$ the set of order statistics in a random sample of size $r$ from $U(0,1)$. We seek transformations of the type ${\cal{T}}: {\bf{U}} \mapsto {\bf{u}}$, that is transformations which from the censored set of order statistics  ${\bf{U}}$ in a sample of size $n$ from $U(0,1)$, lead to a complete set ${\bf{u}}$ of order statistics in a sample of size $r$ from $U(0,1)$. The following transformations have appeared in the literature:
\begin{itemize} \item (1). Michael and Schucany (1979)
\[ u_{i:r}=\frac{U_{i:n}}{U_{r:n}} \left [ B_{r,n-r+1}(U_{r:n})\right]^{1/r}, \ i=1,2 \ldots r, \]
where $B_{r,n-r+1}(u)=\sum_{k=r}^n \frac{n!}{k!(n-k)!}u^k(1-u)^{n-k}$, denotes the DF of the beta distribution with parameters $r$ and $n-r+1$.

\item (2). O'Reilly and Stephens (1988)
\[ u_{i:r}=1- \prod_{j=1}^i \left [\frac{1-U_{j:n}}{1-U_{j-1:n}} \right]^{\frac{n-j+1}{r-j+1}}, \ i=1,2 \ldots r. \]

\item (3). Lin et al. (2008), see also Fischer and Kamps (2011, Theorem 2(3.)). Let
\[ u_{i}= \left[ \frac{1-U_{i:n}}{1-U_{i-1:n}} \right]^{n-i+1}, \ i=1,2 \ldots r. \]
Then set $u_{i:r}=u_{(i)}, \  \ i=1,2 \ldots r$, where $u_{(1)} < u_{(2)}< \ldots < u_{(r)} $ denotes the ordered set of $u_i$.

\item (4). Fischer and Kamps (2011, Theorem 2 (cases 4. and 5.))

\[ u_{i:r}=\prod_{j=i}^r \left [1-\left(\frac{1-U_{j:n}}{1-U_{j-1:n}}\right)^{n-j+1} \right]^{1/j}, \ i=1,2 \ldots r. \]

\item (5). Fischer and Kamps (2011, Theorem 2 (cases 2. and 6.))

\bq
u_{i:r} &=& 1-\left [ 1-B_{r,n-r+1}(U_{r:n})\right]^{1/r}\prod_{j=2}^i \left [1-\left(\frac{U_{r-j+1:n}}{U_{r-j+2:n}}\right)^{r-j+1} \right]^{\frac{1}{r-j+1}}, \\
 &&i=1,2 \ldots r.
\eq

\end{itemize}

As it has already been mentioned, the transformations above are to be combined with a transformation--to--normality.
The aim with this combination is to produce transformed values, say $z_j$, which are stochastically equivalent under the null hypothesis $H_0$ to standardized values which would have been produced in a complete random sample of size $r$ from the standard normal distribution. The latter transformation, which is presented below for the case of a complete sample, was shown to be effective for a wide variety of distributions under testing with uncensored samples; see Meintanis (2009). It has also been applied successfully to the case of testing for the error distribution in generalized linear models by Klar and Meintanis (2011).

\begin{itemize}
\item (6). Chen and Balakrishnan (1995) \\ (i) Efficiently estimate $\vartheta$ by $\widehat \vartheta_n$ based on $X_{j:n}, \ j=1,2 \ldots n$. \\ (ii) Calculate $Y_j=\Phi^{-1}\left({\cal {F}}_{\widehat \vartheta_n}(X_{j:n})\right)$,  $\Phi(\cdot)$ being the standard normal DF. \\ (iii) Compute $Z_j=(Y_j-\bar Y)/s_Y$, where $\bar Y=n^{-1}\sum_{j=1}^n Y_j$, and $s_Y^2=(n-1)^{-1}\sum_{j=1}^n (Y_j- \bar Y)^2$.
\end{itemize}

In the Appendix we provide an analysis of the process produced by the Chen--Bala\-krishnan transformation in an effort to justify the documented validity of this approach under so diversified sampling situations.  There, the process $\hat\beta_{n,2}$ (see (A.1) in the Appendix) is the dominating part and corresponds to testing for normality with estimated parameters, for which efficient (ML) estimators exist, and the test statistics do not depend on the estimates of the parameters nor do they depend on the values of these parameters (mean and standard deviation).

\section{Test statistics}\label{sec_3}

We now illustrate the combined transformation which is suitable for testing the null hypothesis $H_0$ with arbitrary ${\cal {F}}_{\vartheta}$, based on a Type--II censoring scheme.

\vspace{2mm}

\hspace{0.7mm}  {\texttt{TRANSFORMATION (7)}}:

\begin{itemize} \item Efficiently estimate $\vartheta$ by $\widehat \vartheta_r$, based on $X_{j:n}, \ j=1,2 \ldots r$.

\item Calculate $\widehat U_{j:r}={\cal {F}}_{\widehat \vartheta_r}(X_{j:r})$, and set $\widehat {\bf{U}}=(\widehat U_{1:n},\widehat U_{2:n} \ldots \widehat U_{r:n})^{\rm{T}}$

\item Transform to $u_{j:r}={\cal {T}}(\widehat {\bf{U}})$, where ${\cal {T}}$ denotes anyone of the transformations (1)--(5).

\item Replace $n$ by $r$ in transformation (6), and perform step (ii) of this transformation with ${\cal {F}}_{\widehat \vartheta_n}(X_{j:n})$ replaced by $u_{j:r}$.

\item Perform step (iii) of transformation (6), and then apply any test statistic for normality to the values $z_j, \ j=1,2 \ldots r$, so produced.

\end{itemize}

The appropriate normality tests are with estimated parameters and amongst them we consider the classical GOF statistics based on the empirical DF. Specifically,  the Cram\'er--von Mises and the  Anderson--Darling are given by
\be
W^2=\sum_{j=1}^r \left (\Phi(z_j) -\frac{2j-1}{2r}\right)^2 +\frac{1}{12r},
\ee
and
\be
A^2=-r-\frac{1}{r}\sum_{j=1}^r \left [(2j-1) \log \Phi(z_j) +(2r+1-2j) \log (1-\Phi(z_j)) \right],
\ee
respectively. Asymptotic percentage points and modifications of the statistics for finite sample size can be found
in Table 4.7 in D'Agostino and Stephens (1986).

We also consider a test for normality which utilizes the characteristic function (CF) and takes the form
\be
C^2= r \int_{-\infty}^\infty |\widehat \varphi_r(t)-e^{-(1/2)t^2}|^2 \ w(t) dt,\ee
where $\widehat \varphi_r(t)=r^{-1} \sum_{j=1}^r e^{it z_j}$ is the empirical CF  of $z_j, \ j=1,...,r$, and $w(t)$ denotes a weight function introduced in order to smooth out the periodic behavior of $\widehat \varphi_r(t)$. Note that the test statistic $C^2$ compares the  empirical CF of $z_j$ to the CF of the standard normal distribution. For $w(t)=e^{-a t^2}, \ a>0$, we have from (3.3) after some straightforward algebra,
\be
C^2 := C_a^2 &=& \frac{1}{r} \sqrt{\frac{\pi}{a}}\sum_{j,k=1}^r e^{-(z_j-z_k)^2/4a} \\
    && - 2 \sqrt{\frac{2\pi}{1+2a}}\sum_{j=1}^r e^{-z_j^2/(2+4a)} + r \sqrt{\frac{\pi}{1+a}} \ .
\ee
Epps and Pulley (1983) proposed this test statistic and showed that $C_a^2$ is very competitive to the classical tests $W^2$ and $A^2$. Despite the fact that the asymptotic null distribution of this statistic is complicated, there exist some approximations thereof; see for instance Henze (1990) for an approximation based on Johnson distributions. In fact, by using a simple transformation of $C_a^2$, the  test can be easily carried out for finite samples provided that the sample size is larger than or equal to 10 (Henze, 1990, p. 17).

In connection with the weight function we point out that the choice $w(t)=e^{-a t^2}$ has become something of a standard for the CF statistic in (3.3); see for instance Epps and Pulley (1983), Epps (2005) and Henze and Wagner (1997). Other weight functions are also possible, for instance  $w(t)=e^{-a |t|}$, but it is well known that the specific functional form of $w(t)$ is not so important. This conclusion is based on the equivalence of the CF statistic to an L2 distance--statistic involving density estimators (see Bowman and Foster, 1993), and the corresponding association of  the weight fuction to the kernel function in density estimation. On the other hand, the value of the weight parameter $a$ has a greater impact on the power properties of the CF statistic as it has been related to the choice of the bandwidth in density estimation. The only analytic treatment available on objective optimal values of $a$ is provided in Tenreiro (2009) by relating this value to the local Bahadur slopes of the test statistic. Even with these analytical results, specific quantitative suggestions require consideration of specific deviations from the null hypothesis of normality. Nevertheless Tenreiro (2009) recovers what was already a common practice in simulations, namely that smaller values (resp. larger values) of $a$ are appropriate for detecting short--tailed (resp. long--tailed) alternatives. Based on these theoretical considerations as well as on extensive simulations, he suggests a bandwith of 0.71, corresponding to $a=0.5$, as an overall compromise choice. This conclusion agrees with the results described in Epps and Pulley (1983); the same weight  was also chosen in other studies like those of Baringhaus et al. (1989) and Arcones and Wang (2006). As a consequence, we also employed the weight $a=0.5$ in our simulations.

We close this section by noting that performing a test for uniformity after the second step of transformation (7), which would in fact seem as a reasonable and simpler approach,  does not lead to a valid testing procedure since the test statistics would then depend on the parameter estimate and the parameter value. We will take up and further clarify this point again in the Appendix. Also note that Chen and Balakrishnan (1995) suggested transformation (6) (and provided partial justification for), in the case of  the tests $W^2$ and $A^2$. Nevertheless, the Kolmogorov--Smirnov statistic would have also been a reasonable competitor in this study, however we opt not to included it as it is generally known to be less powerful compared to the Cram\'er--von Mises and the  Anderson--Darling tests; see for instance Castro--Kuriss (2011) or Glen and Foote (2009).

\section{Estimation of parameters}\label{sec_4}

In order to implement transformation (7) we require an efficient estimator $\widehat \vartheta_r$ of the parameter $\vartheta$, such as the maximum likelihood estimator (MLE). This estimator employs the censored data $X_{j:r}, \ j=1,2 \ldots r$, and will depend on the specific parametric form of ${\cal{F}}_\vartheta$ under the null hypothesis $H_0$. The following parametric distributions are of special interest:

\begin{itemize}
\item The exponential distribution $Exp(\sigma)$ with DF, $F(x)=1-e^{-x/\sigma}$. Then the MLE is given by
\[
\widehat \sigma=\frac{\sum_{j=1}^r X_{j:n}+(n-r)X_{r:n}}{r}.
\]

\item The gamma distribution $\gamma(\theta,\sigma)$ with density, $(\sigma^\theta \Gamma(\theta))^{-1}x^{\theta-1} e^{-x/\sigma}$. A simplified form of the MLE equations is given by (see Wilk et al., 1962 or Johnson et al., 1994),
\[
r \log {\cal {P}}_r=n \left[ \frac{\Gamma'(\widehat \theta)}{\Gamma(\widehat \theta)}-\log \frac{X_{r:n}}{\widehat \sigma}\right]-(n-r) \frac{\partial \log J(\widehat \theta)}{\partial\widehat \theta},
\]
\[
\frac{X_{r:n}{\cal {S}}_r}{\widehat \sigma}= \widehat \theta- \frac{(n-r)e^{-X_{r:n}/\widehat \sigma}} {rJ(\widehat \theta)} ,
\]
where $J(\widehat\theta):=J(\widehat \theta,X_{r:n}/\widehat \sigma)$, with $J(x,y)=\int_1^\infty t^{x-1} e^{-y t}dt$, and
\[
{\cal {P}}_r=\frac{\left(\prod_{j=1}^r X_{j:n}\right)^{1/r}}{X_{r:n}}, \ \
{\cal {S}}_r=\frac{\sum_{j=1}^r X_{j:n}}{rX_{r:n}}.
\]

\item The normal distribution $\mathcal{N}(\mu,\sigma^2)$ with mean $\mu$ and variance $\sigma^2$. We employ the estimates \[ \widehat \mu=\sum_{j=1}^r b_j X_{j:r}, \ \ \widehat \sigma=\sum_{j=1}^r c_j X_{j:r},
\] suggested by Gupta (1952). This author provided the values of the coefficients $(b_j,c_j)$ for $n \leq 10$. For larger sample sizes the values suggested are \[ b_j=\frac{1}{r}-\frac{\bar m (m_j-\bar m)}{\sum_{j=1}^r (m_j-\bar m)^2}, \ \ c_j=\frac{m_j-\bar m}{\sum_{j=1}^r (m_j-\bar m)^2}, \] where $m_j$ denotes the expected value of the $j^{\rm{th}}$ order statistic in a sample of size $n$ from the standard normal distribution, and $\bar m=r^{-1}\sum_{j=1}^r m_j$.
There is also an  approximate method whereby $m_j$ is replaced by $\Phi^{-1}((j-0.375)/(n+0.125))$; see D'Agostino and Stephens (1986).

We have also implemented as an alternative estimation method the modified maximum likelihood estimation proposed by Tiku and co--workers. This method uses a suitable linearization of the likelihood function; see Tiku (1967) or Tiku, Tan, and Balakrishnan (1986).
\end{itemize}


\section{Simulations}\label{sec_5}

Tables \ref{Allgemeines_Verfahren_(exp_n=100) (Teil 1)} to \ref{Allgemeines_Verfahren_(norm_n=100) (Teil 2)}
show parts of the results of extensive simulation studies with the testing procedures presented in Section \ref{sec_3}.
Specifically we employ the Cram\'er-von Mises test $W^2$, the Anderson-Darling test $A^2$, and the characteristic function test
with weight function $e^{-t^2/2}$, denoted by $C^2$. 
We used transformations (1) to (5) (see Section 2), abbreviated as MS, OS, LHB, FK1 and FK2.

As hypothetical models, the exponential, gamma and normal distribution have been used,
with estimation of parameters carried out by the methods outlined in Section 4.

\begin{sloppypar}
As alternatives we employed the following distributions which are often used in lifetime and failure analysis:
Weibull distribution Wei$(\alpha,\beta)$ with density
$\frac{\alpha}{\beta}\big(\frac{x}{\beta}\big)^{\alpha-1} \exp\big(-(\frac{x}{\beta})^\alpha\big)$,
inverse Gaussian distribution $IG(\mu,\lambda) $ with density
$\big(\frac{\lambda}{2\pi x^3}\big)^{1/2} \exp\big(\frac{-\lambda(x-\mu)^2}{2\mu^2x}\big)$,
logarithmic gamma distribution $\mathcal{L}\gamma(\alpha,\beta)$ with density
$\frac{\beta^\alpha}{\Gamma(\alpha)x^{\beta+1}} (\log x)^{\alpha-1}$,
logistic distribution $\mathcal{L}(\alpha,\beta)$ with density
$\frac1\beta (1+\exp(-\frac{x-\alpha}{\beta}))^{-2} \exp(-\frac{x-\alpha}{\beta})$,
lognormal distribution $\mathcal{LN}(\mu,\sigma)$ with density
$(\sqrt{2\pi}\sigma x)^{-1} \exp(\frac{-(\log x-\mu)^2}{2\sigma^2})$,
Student's $t$-distribution with $m$ degrees of freedom $t_m$,
and the three distributions which were used as hypothetical models.
\end{sloppypar}

The censoring proportions considered are $50\%$ and $25\%$, which corresponds to $r/n=0.50$ and $0.75$, respectively, with sample sizes $n=40$ and $n=100$.
The entries in the tables give the percentage of rejection of the respective hypothesis based on $10~000$ repetitions, at nominal level of significance $\alpha=0.05$. A rejection rate of $100\%$ is indicated by $\ast$.

All tests have been also performed at nominal levels of significance $\alpha=0.01$ and $\alpha=0.1$.
Since however the relative performance was unchanged, the corresponding results are omitted.
Likewise, the results for the (very high) censoring proportion of $75\%$ are omitted, but some of the following remarks also apply to this case.

All simulations have been done using the statistical computing environment R (R Core Team, 2012).

\subsection{Testing for exponentiality}

\begin{figure}[htp]
\subfigure[$\gamma(4,1)$]{\includegraphics[width=0.33\textwidth]{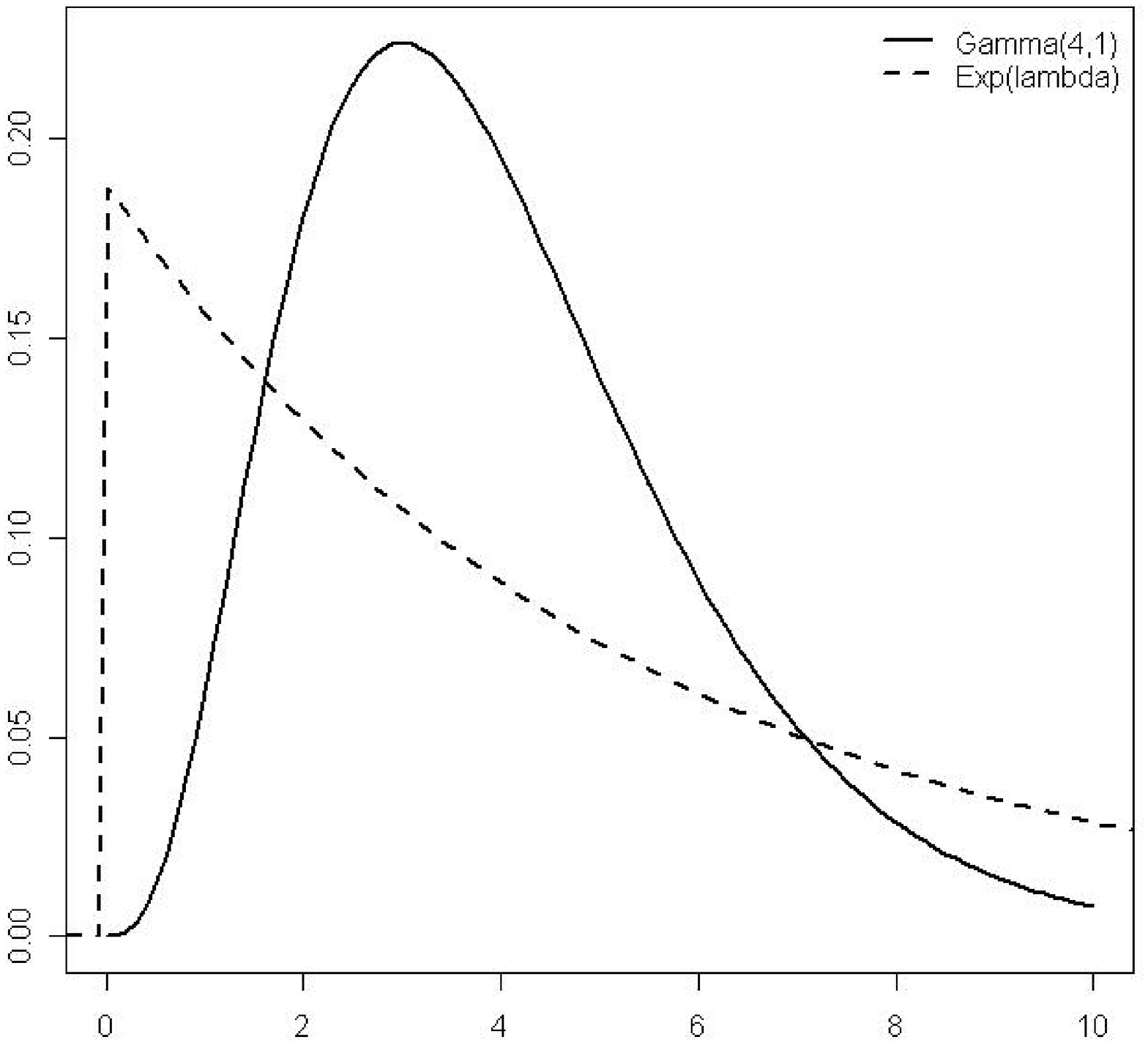}}\hfill
\subfigure[$\mathcal{L}\gamma(2,1)$]{\includegraphics[width=0.33\textwidth]{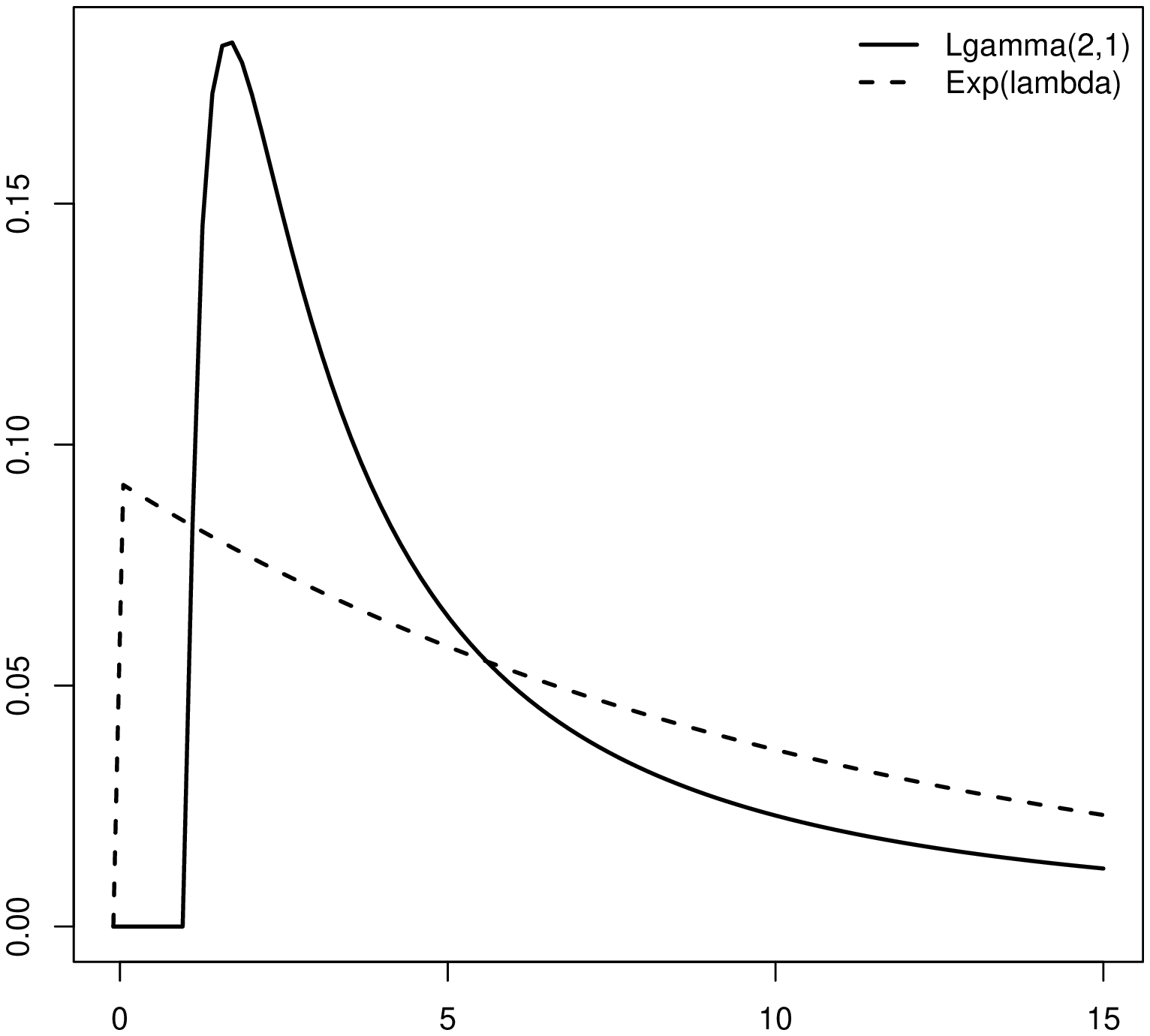}}\hfill
\subfigure[$Wei(2,1)$]{\includegraphics[width=0.33\textwidth]{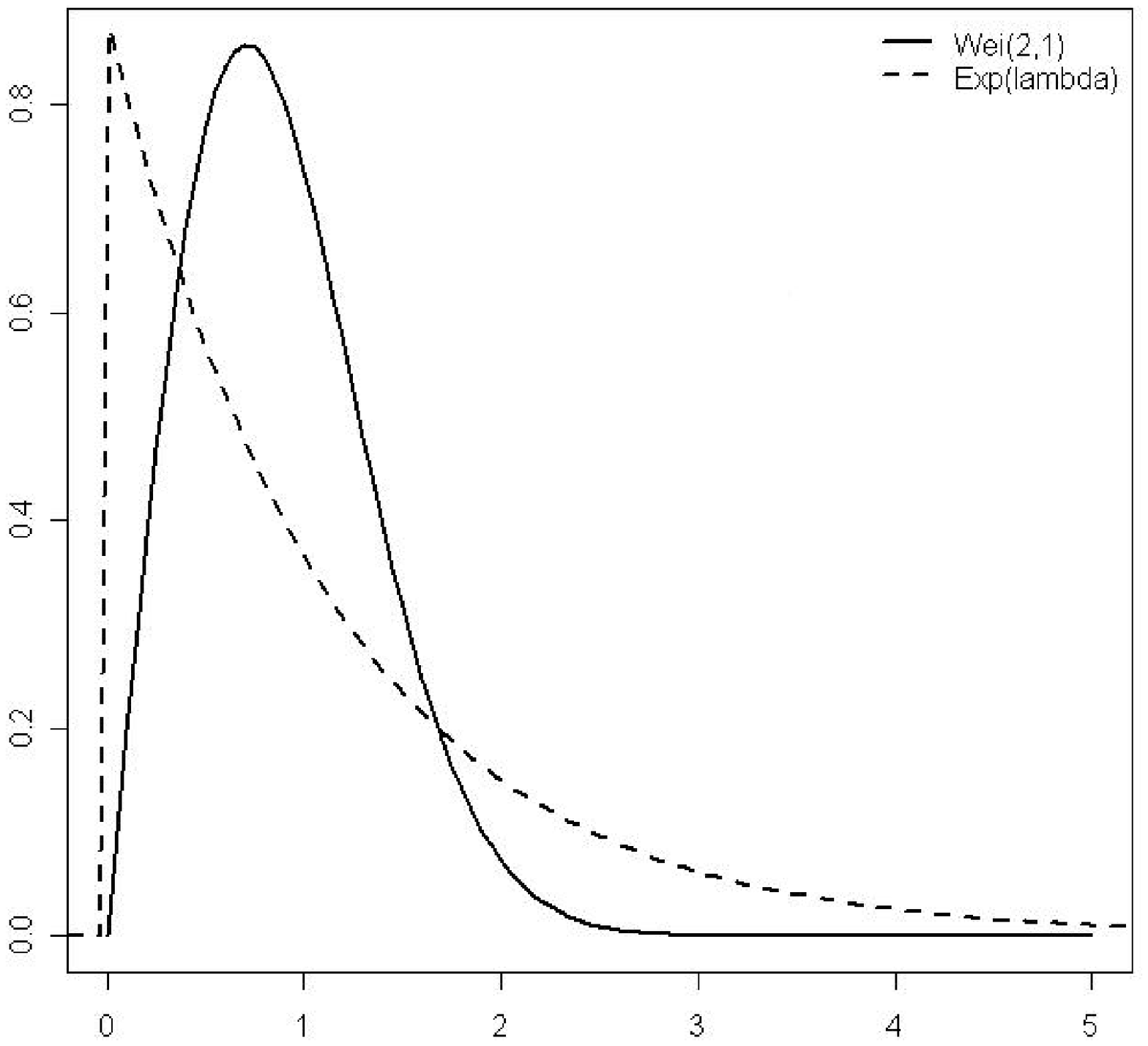}}\hfill
\subfigure[$IG(4,1)$ ]{\includegraphics[width=0.33\textwidth]{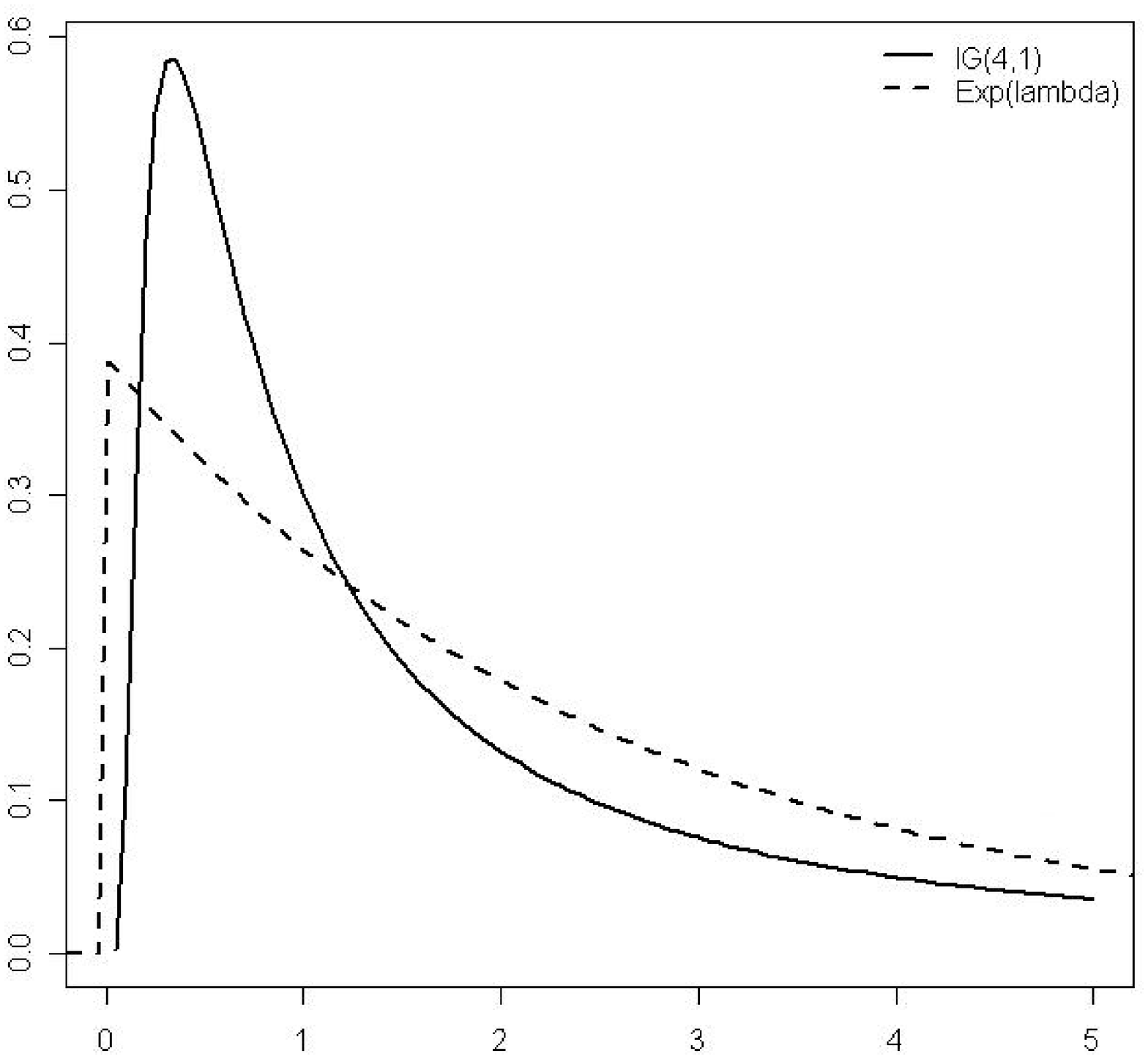}}\hfill
\subfigure[$IG(1,4)$ ]{\includegraphics[width=0.33\textwidth]{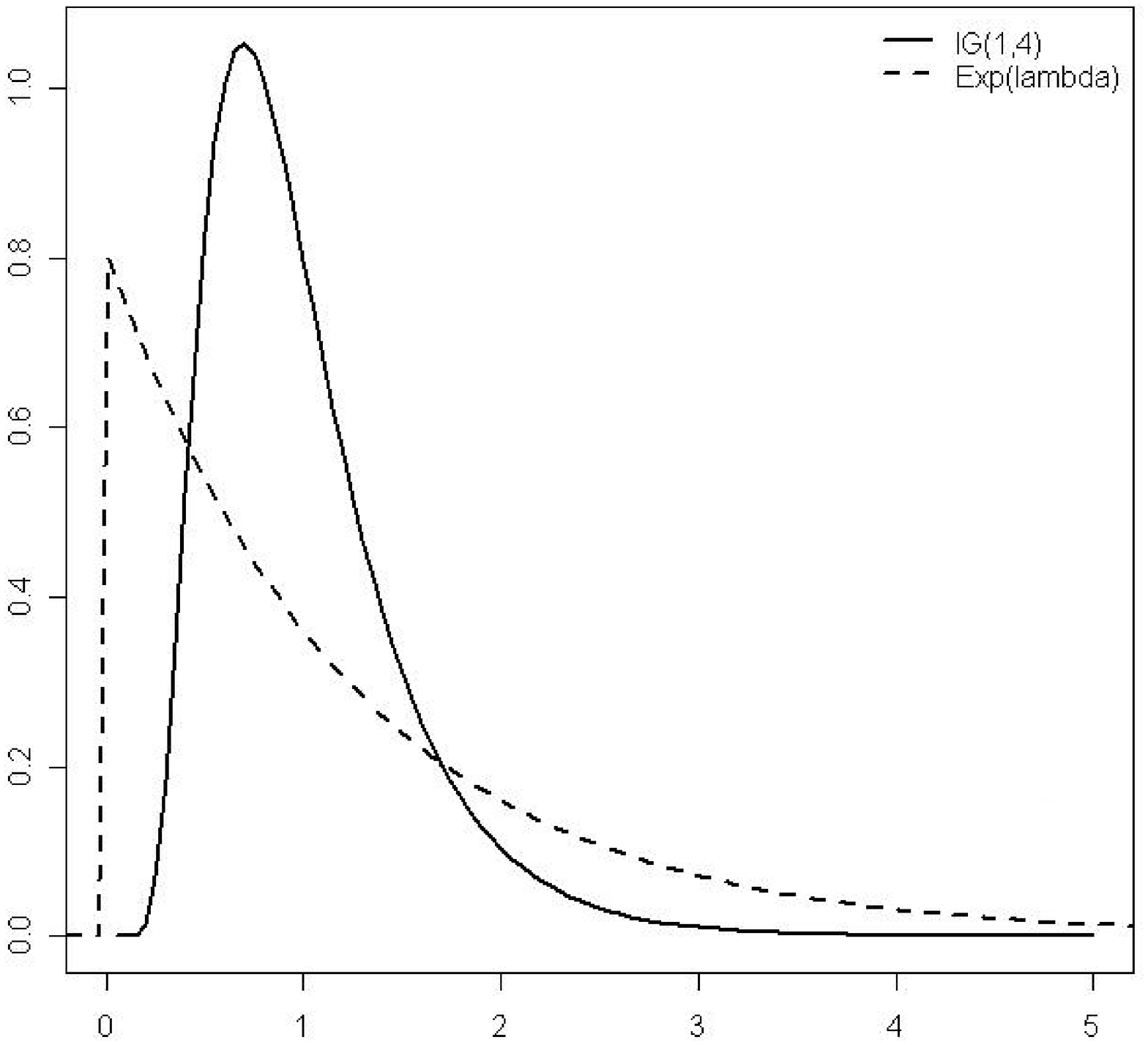}}\hfill
\subfigure[$\mathcal{LN}(0,1)$]{\includegraphics[width=0.33\textwidth]{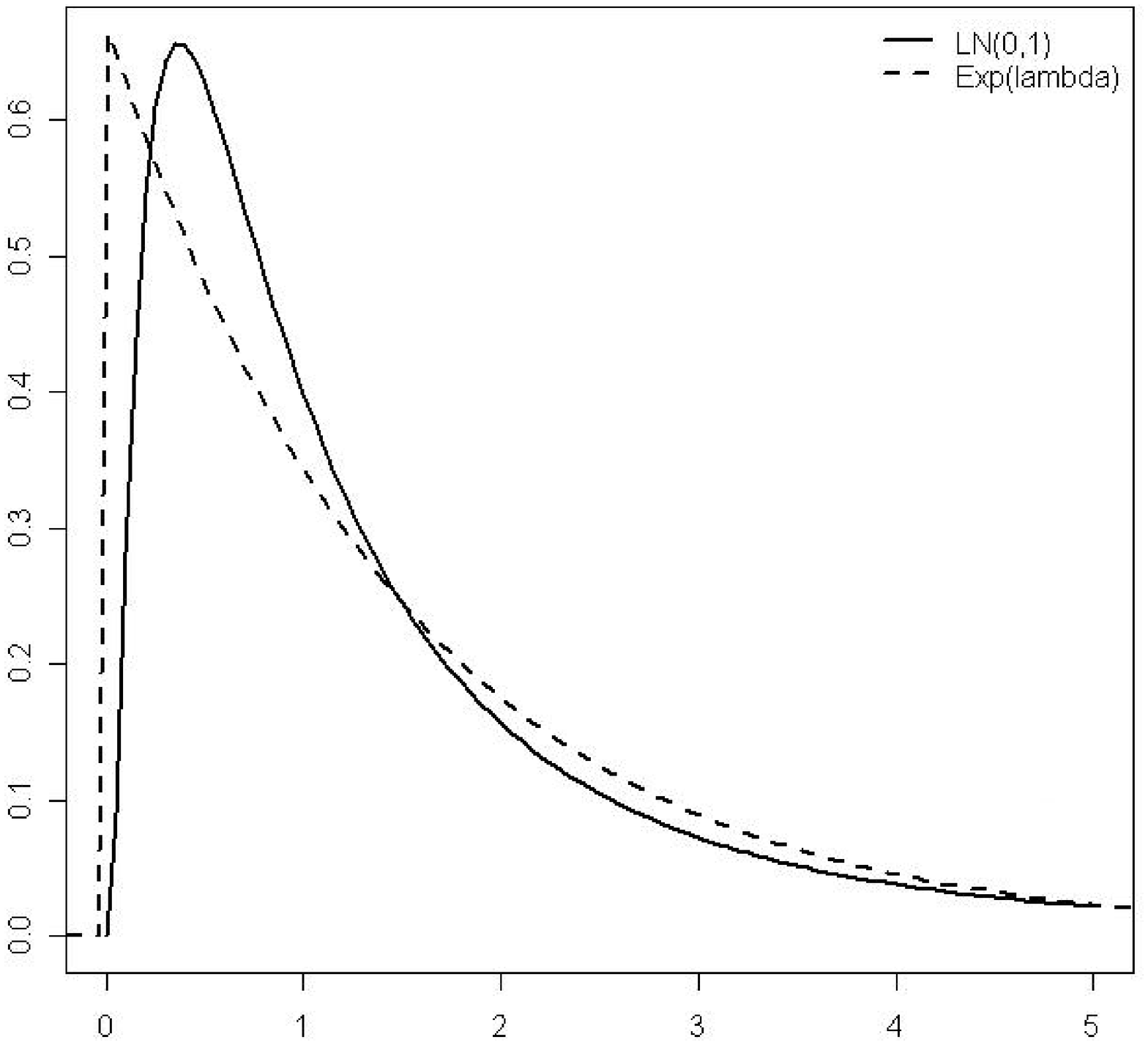}}\hfill
\caption{Densities of some alternative distributions compared with the density of a fitted exponential distribution.
Parameter estimation is based on a sample from the corresponding alternative with $n=100$ and $r=75$.}
\label{Vergleich mit Exponentialverteilung}
\end{figure}

Simulation results for testing the hypothesis of exponentiality are given in Tables
\ref{Allgemeines_Verfahren_(exp_n=100) (Teil 1)} and \ref{Allgemeines_Verfahren_(exp_n=100) (Teil 2)}. The conclusions drawn from these results are as follows:

\begin{description}
\item[Level:] Most procedures maintain the nominal level very well, with the tests based on the MS and FK2 transforms being somewhat conservative.

\item[\boldmath $\gamma(\alpha,\beta)$ and $\mathcal{L}\gamma(\alpha,\beta)$ alternatives:] (Figure \ref{Vergleich mit Exponentialverteilung}(a)(b))
Due to the similarity of the gamma to the exponential distribution, detecting this hypothesis is difficult. In fact, transformations MS, OS and FK2 seem completely unsuitable for this purpose. On the other hand, the tests based on the LHB transformation seem to work best against gamma alternatives, while against logarithmic gamma alternatives, the tests based on MS and OS are also competitive.

For the $\mathcal{L}\gamma(2,1)$ distribution, all tests based on the LHB transformation show an astonishing behaviour:
The power is sharply decreasing when $r$ is increasing, i.e. with a lower degree of censoring.
The same can be seen for further alternatives.
This behaviour of the LHB transform can also be observed when testing a simple hypotheses,
see Tables I-IV in Lin et al. (2008): there, the power of the Anderson-Darling test combined with the LHB transform,
called ${}_{\stackrel{\ast}{T}} A_r^2$, decreases with $r$ for several alternatives (in particular for alternatives
$F_{41}$ and $F_{51}$ which are defined on p. 634 in Lin et al. (2008)).
Similar patterns also occur with other transformations, see the results for the Anderson-Darling test
combined with the MS transform, called ${}_{T} A_r^2$, in Lin et al. (2008), and Table 1 in O'Reilly and Stephens (1988)
for tests based on the OS transformation.

\item[\boldmath $Wei(\alpha,\beta)$ alternatives:] (Figure \ref{Vergleich mit Exponentialverteilung}(c))
Again, tests based on the LHB transform are the best, followed by FK1, while the tests based on the  FK2 transformation is clearly unable to detect the Weibull alternatives
used in the simulations.

\item[\boldmath $IG(\mu,\lambda)$ and $\mathcal{LN}(\mu,\sigma)$ alternatives:] (Figure \ref{Vergleich mit Exponentialverteilung}(d)(e)(f))
For these distributions, the tests based on FK1 and FK2 do not work.
Interestingly, the MS and OS--based tests work much better against $IG(4,1)$ than LHB, while the converse holds true for the
$IG(1,4)$ distribution.
The power results against the $\mathcal{LN}(0,1)$ distribution comply with the fact that this distribution is hard to distinguish from a suitable exponential distribution  (see Figure \ref{Vergleich mit Exponentialverteilung}(d)).

\item[Summary:]
The tests based on the transformation of Lin, Huang and Balakrishnan should be used,
while the tests based on FK1 and FK2 do not work well. Also within one transformation, the difference between the four test statistics is not always particularly noticeable. Nevertheless, in many cases the characteristic function based test has slightly higher power than the tests based on the empirical distribution function.

In the simulations for the exponential distribution, we added results for the two most common direct statistics (DS), i.e., statistics which are modifications of corresponding full--sample versions and may be applied directly to the original censored data, without transformations. These tests are the Cram\'er-von Mises and the Anderson-Darling test; see Section 4.9.5 of D'Agostino, Stephens (1986).
Corresponding results are given in obvious notation in the last two columns in Tables \ref{Allgemeines_Verfahren_(exp_n=100) (Teil 1)} and \ref{Allgemeines_Verfahren_(exp_n=100) (Teil 2)}.
Generally, the newly proposed tests have inferior power. Recall however their advantage of general applicability: We do not need new critical values for each distribution, sample size and censoring proportion, not to mention dependence on parameters and corresponding estimators. In this connection note that since even for this standard distribution critical values for our censoring proportions 25\% and 50\% are not available in the literature, they are provided in Table \ref{Critical_values_exp_censored_Data}.

\end{description}

\renewcommand{\baselinestretch}{1.0}
\setlength{\tabcolsep}{2mm}
\begin{table}[htp]
\begin{center}

\begin{tabular}{cc||c|c|c|c|c|c}
  \hline
 \multirow{2}{*}{$n$} &\multirow{2}{*}{$\frac{r}{n}$} & \multicolumn{3}{c|}{$A^2_{r,n}$} & \multicolumn{3}{c}{$W^2_{r,n}$}\\
 & & $10\%$ & $5\%$ & \multicolumn{1}{c|}{$1\%$} & $10\%$ & $5\%$ & \multicolumn{1}{c}{$1\%$}\\
  \hline\hline	
 40 & 0.50 & 0.479 & 0.609 & 0.932 & 0.063 & 0.081 & 0.124 \\
 40 & 0.75 & 0.726 & 0.918 & 1.401 & 0.12 & 0.152 & 0.232 \\ \hline
100 & 0.50 & 0.483& 0.617& 0.953& 0.063& 0.082& 0.127\\
100 & 0.75 & 0.734& 0.930& 1.414& 0.121& 0.154& 0.237\\ \hline
\end{tabular}

\end{center}
\caption{Critical values of direct edf statistics for testing for exponentiality based on $10^7$ replications}
\label{Critical_values_exp_censored_Data}
\end{table}

\setlength{\tabcolsep}{1mm}
\begin{landscape}
\begin{table}
\begin{center}

\begin{tabular}{ccc|ccc|ccc|ccc|ccc|ccc|cc}
  \hline
 \multirow{2}{*}{Distribution} & \multirow{2}{*}{$\frac{r}{n}*100\%$} & \multirow{2}{*}{$n$} & \multicolumn{3}{c|}{MS} & \multicolumn{3}{c|}{OS} & \multicolumn{3}{c|}{LHB} & \multicolumn{3}{c|}{FK1} & \multicolumn{3}{c|}{FK2} & \multicolumn{2}{c}{DS} \\
 & & & $A^2$ & $W^2$ & \multicolumn{1}{c|}{$C^2$} & $A^2$ & $W^2$ & \multicolumn{1}{c|}{$C^2$}
     & $A^2$ & $W^2$ & \multicolumn{1}{c|}{$C^2$} & $A^2$ & $W^2$ & \multicolumn{1}{c|}{$C^2$}
     & $A^2$ & $W^2$ & \multicolumn{1}{c|}{$C^2$} &$A^2_{r,n}$ & $W^2_{r,n}$ \\
  \hline\hline
\multirow{4}{*}{$Exp(1)$}
  & 50 & 40 & 4 & 4 & 3 & 6 & 6 & 5 & 5 & 5 & 4 & 5 & 5 & 4 & 4 & 4 & 3 & 5 & 5 \\
  & 50 &100 & 4 & 4 & 3 & 5 & 5 & 5 & 5 & 5 & 5 & 5 & 5 & 4 & 4 & 4 & 3 & 4 & 4 \\ \cline{2-20}
  & 75 & 40 & 4 & 4 & 4 & 5 & 5 & 5 & 5 & 5 & 5 & 5 & 5 & 5 & 4 & 4 & 4 & 5 & 5 \\
  & 75 &100 & 4 & 5 & 4 & 6 & 5 & 6 & 5 & 5 & 5 & 5 & 5 & 5 & 4 & 4 & 4 & 5 & 5 \\ \hline
\multirow{4}{*}{$\gamma(2,1)$}
  & 50 & 40 & 4 & 4 & 3 & 6 & 6 & 6 & 7 & 6 & 7 & 7 & 6 & 6 & 0 & 0 & 0 & 53 & 53 \\
  & 50 &100 & 4 & 4 & 4 & 6 & 6 & 6 &12 &11 &15 &12 & 9 &14 & 0 & 0 & 1 & 95 & 92 \\ \cline{2-20}
  & 75 & 40 & 5 & 4 & 4 & 6 & 5 & 6 & 8 & 8 & 9 & 8 & 7 & 8 & 0 & 0 & 0 & 70 & 68 \\
  & 75 &100 & 6 & 5 & 5 & 6 & 6 & 6 &17 &14 &20 &16 &13 &19 & 0 & 1 & 1 & 99 & 98 \\ \hline
\multirow{4}{*}{$\gamma(4,1)$}
  & 50 & 40 & 5 & 5 & 3 & 7 & 6 & 6 &28 &24 &33 & 7 & 6 & 7 & 0 & 0 & 0 & 99 &  99 \\
  & 50 &100 & 6 & 5 & 4 & 6 & 6 & 6 &74 &63 &80 &15 &10 &18 & 0 & 0 & 0 &100 & 100 \\ \cline{2-20}
  & 75 & 40 & 5 & 5 & 4 & 6 & 6 & 6 &42 &34 &49 & 9 & 7 &10 & 0 & 0 & 0 &100 & 100 \\
  & 75 &100 & 7 & 6 & 6 & 5 & 5 & 5 &89 &80 &93 &23 &16 &29 & 0 & 0 & 0 &100 & 100 \\ \hline
\multirow{4}{*}{$IG(4,1)$}
  & 50 & 40 & 15 & 14 & 15 & 13 & 12 & 11 &  5 &  5 &  5 & 4 & 3 & 4 &  2 &  2 &  2 & 19 & 16 \\
  & 50 &100 & 36 & 32 & 41 & 38 & 34 & 42 & 10 &  9 & 12 & 5 & 4 & 4 &  4 &  5 &  6 & 62 & 37 \\ \cline{2-20}
  & 75 & 40 & 39 & 35 & 43 & 36 & 33 & 39 &  6 &  6 &  6 & 5 & 4 & 5 & 15 & 14 & 17 & 23 & 26 \\
  & 75 &100 & 79 & 72 & 83 & 86 & 80 & 89 & 13 & 12 & 15 & 7 & 5 & 6 & 36 & 32 & 44 & 64 & 58 \\ \hline
\multirow{4}{*}{$IG(1,4)$}
  & 50 & 40 &  5 &  5 &  3 &  6 &  6 &  5 & 59 & 48 & 67 &  3 & 2 &  3 & 0 & 0 & 0 & 100 & 100 \\
  & 50 &100 & 11 & 10 &  9 & 10 &  9 & 10 & 99 & 94 & 99 &  7 & 5 & 10 & 0 & 0 & 0 & 100 & 100 \\ \cline{2-20}
  & 75 & 40 &  8 &  8 &  6 &  8 &  7 &  7 & 74 & 61 & 82 &  5 & 3 &  6 & 0 & 0 & 0 & 100 & 100 \\
  & 75 &100 & 18 & 16 & 18 & 16 & 15 & 18 &100 & 99 &100 & 12 & 7 & 17 & 0 & 0 & 1 & 100 & 100 \\ \hline
\hline
\end{tabular}

\end{center}
\caption{Percentage of rejection of tests for exponentiality based on 10000 replications (part 1)}
\label{Allgemeines_Verfahren_(exp_n=100) (Teil 1)}
\end{table}
\end{landscape}

\begin{landscape}
\begin{table}
\begin{center}

\begin{tabular}{ccc|ccc|ccc|ccc|ccc|ccc|cc}
  \hline
 \multirow{2}{*}{Distribution} & \multirow{2}{*}{$\frac{r}{n}*100\%$} & \multirow{2}{*}{$n$} & \multicolumn{3}{c|}{MS} & \multicolumn{3}{c|}{OS} & \multicolumn{3}{c|}{LHB} & \multicolumn{3}{c|}{FK1} & \multicolumn{3}{c|}{FK2} & \multicolumn{2}{c}{DS} \\
 & & & $A^2$ & $W^2$ & \multicolumn{1}{c|}{$C^2$} & $A^2$ & $W^2$ & \multicolumn{1}{c|}{$C^2$}
     & $A^2$ & $W^2$ & \multicolumn{1}{c|}{$C^2$} & $A^2$ & $W^2$ & \multicolumn{1}{c|}{$C^2$}
     & $A^2$ & $W^2$ & \multicolumn{1}{c|}{$C^2$} &$A^2_{r,n}$ & $W^2_{r,n}$ \\
  \hline\hline
  \multirow{4}{*}{$Wei(2,1)$}
  & 50 & 40 & 5 & 5 & 4 & 10 &  9 & 10 & 12 & 11 & 14 &  8 &  7 &  8 & 0 & 0 & 0 & 84 & 85 \\
  & 50 &100 & 5 & 5 & 5 & 12 & 11 & 13 & 30 & 25 & 36 & 19 & 14 & 21 & 0 & 0 & 0 &100 &100 \\ \cline{2-20}
  & 75 & 40 & 8 & 7 & 7 & 12 & 11 & 13 & 20 & 17 & 23 & 12 & 10 & 13 & 0 & 0 & 0 & 98 & 98 \\
  & 75 &100 &10 & 8 &11 & 17 & 15 & 21 & 50 & 41 & 56 & 30 & 22 & 35 & 0 & 0 & 0 &100 &100 \\ \hline
  \multirow{4}{*}{$Wei(4,1)$}
  & 50 & 40 & 11 & 10 & 10 & 14 & 13 & 15 & 65 & 57 & 72 &  7 &  5 &  7 & 0 & 0 & 0 & 100 & 100 \\
  & 50 &100 & 15 & 12 & 14 & 18 & 16 & 21 & 99 & 97 & 99 & 21 & 14 & 27 & 0 & 0 & 0 & 100 & 100 \\ \cline{2-20}
  & 75 & 40 & 20 & 17 & 20 & 19 & 17 & 21 & 88 & 81 & 92 & 11 &  8 & 14 & 0 & 0 & 0 & 100 & 100 \\
  & 75 &100 & 41 & 31 & 43 & 31 & 28 & 36 &100 &100 &100 & 40 & 28 & 51 & 0 & 0 & 0 & 100 & 100 \\ \hline
  \multirow{4}{*}{$\mathcal{L}\gamma(2,1)$}
  & 50 & 40 & 30 & 27 & 31 & 29 & 26 & 28 & 19 & 16 & 23 & 2 & 2 & 2 &  1 &  1 &  2 & 73 & 61 \\
  & 50 &100 & 75 & 67 & 77 & 80 & 73 & 82 & 73 & 52 & 80 & 2 & 1 & 2 &  5 &  6 & 12 &100 & 99 \\ \cline{2-20}
  & 75 & 40 & 71 & 65 & 73 & 70 & 64 & 71 & 16 & 14 & 19 & 3 & 2 & 3 & 20 & 19 & 26 & 55 & 46 \\
  & 75 &100 & 99 & 97 & 99 &100 & 99 & 99 & 50 & 40 & 57 & 5 & 4 & 6 & 57 & 51 & 69 &100 & 93 \\ \hline
  \multirow{4}{*}{$\mathcal{L}\gamma(4,1)$}
  & 50 & 40 & 22 & 20 & 24 & 19 & 17 & 18 &  6 &  5 &  5 &  4 & 3 &  3 &  8 &  8 &  8 & 13 & 15 \\
  & 50 &100 & 51 & 45 & 57 & 58 & 51 & 63 &  8 &  8 &  9 &  6 & 5 &  5 & 16 & 16 & 20 & 32 & 33 \\ \cline{2-20}
  & 75 & 40 & 58 & 51 & 63 & 57 & 52 & 60 & 19 & 17 & 20 &  5 & 4 &  5 & 48 & 44 & 47 & 76 & 78 \\
  & 75 &100 & 93 & 87 & 94 & 97 & 95 & 98 & 37 & 34 & 42 & 11 & 9 & 11 & 82 & 76 & 84 & 98 & 99 \\ \hline
  \multirow{4}{*}{$\mathcal{LN}(0,1)$}
  & 50 & 40 &  6 &  6 &  5 &  6 &  6 &  4 & 5 & 5 &  5 & 6 & 4 & 5 & 0 & 1 & 1 & 34 & 32 \\
  & 50 &100 & 12 & 11 & 12 & 10 & 10 & 10 &10 & 8 & 12 & 7 & 5 & 8 & 1 & 1 & 2 & 84 & 70 \\ \cline{2-20}
  & 75 & 40 & 11 & 10 & 12 &  9 &  9 &  9 & 5 & 5 &  5 & 6 & 5 & 6 & 2 & 2 & 2 & 26 & 22 \\
  & 75 &100 & 26 & 23 & 30 & 25 & 23 & 28 & 8 & 8 & 10 & 7 & 5 & 7 & 3 & 4 & 5 & 73 & 53 \\ \hline
   \hline
\end{tabular}

\end{center}
\caption{Percentage of rejection of tests for exponentiality based on 10000 replications (part 2)}
\label{Allgemeines_Verfahren_(exp_n=100) (Teil 2)}
\end{table}
\end{landscape}

\renewcommand{\baselinestretch}{1.2}


\subsection{Testing for gamma distribution}

Simulation results for testing the gamma hypothesis are given in Tables
\ref{Allgemeines_Verfahren_(gamma_n=100) (Teil 1)} and \ref{Allgemeines_Verfahren_(gamma_n=100) (Teil 2)}. The conclusions drawn from these results are as follows:

\begin{figure}[htp]
\subfigure[$\mathcal{L}\gamma(2,1)$]{\includegraphics[width=0.33\textwidth]{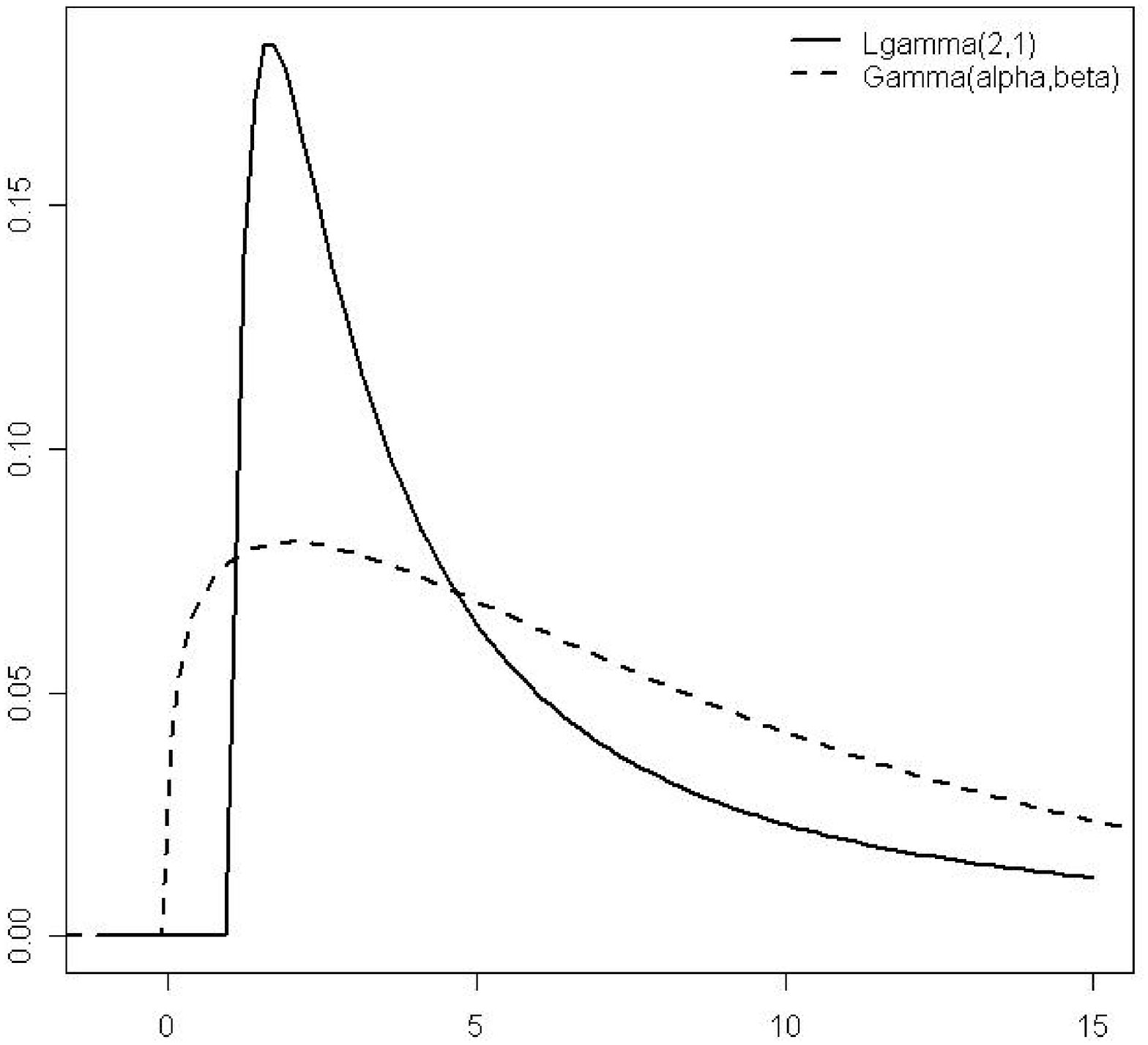}}\hfill
\subfigure[$\mathcal{L}\gamma(4,1)$]{\includegraphics[width=0.33\textwidth]{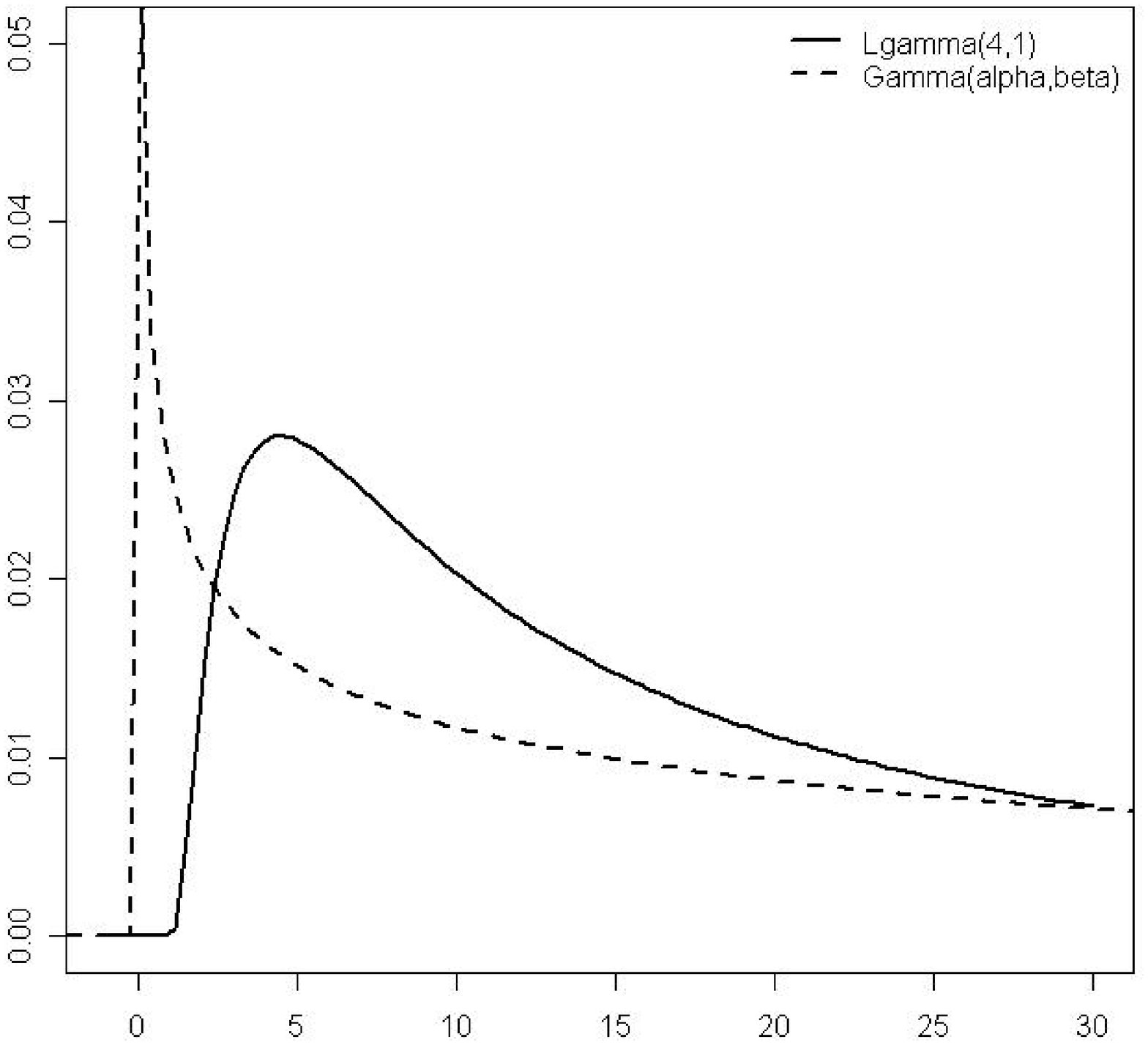}}\hfill
\subfigure[$\mathcal{N}(3,1)$]{\includegraphics[width=0.33\textwidth]{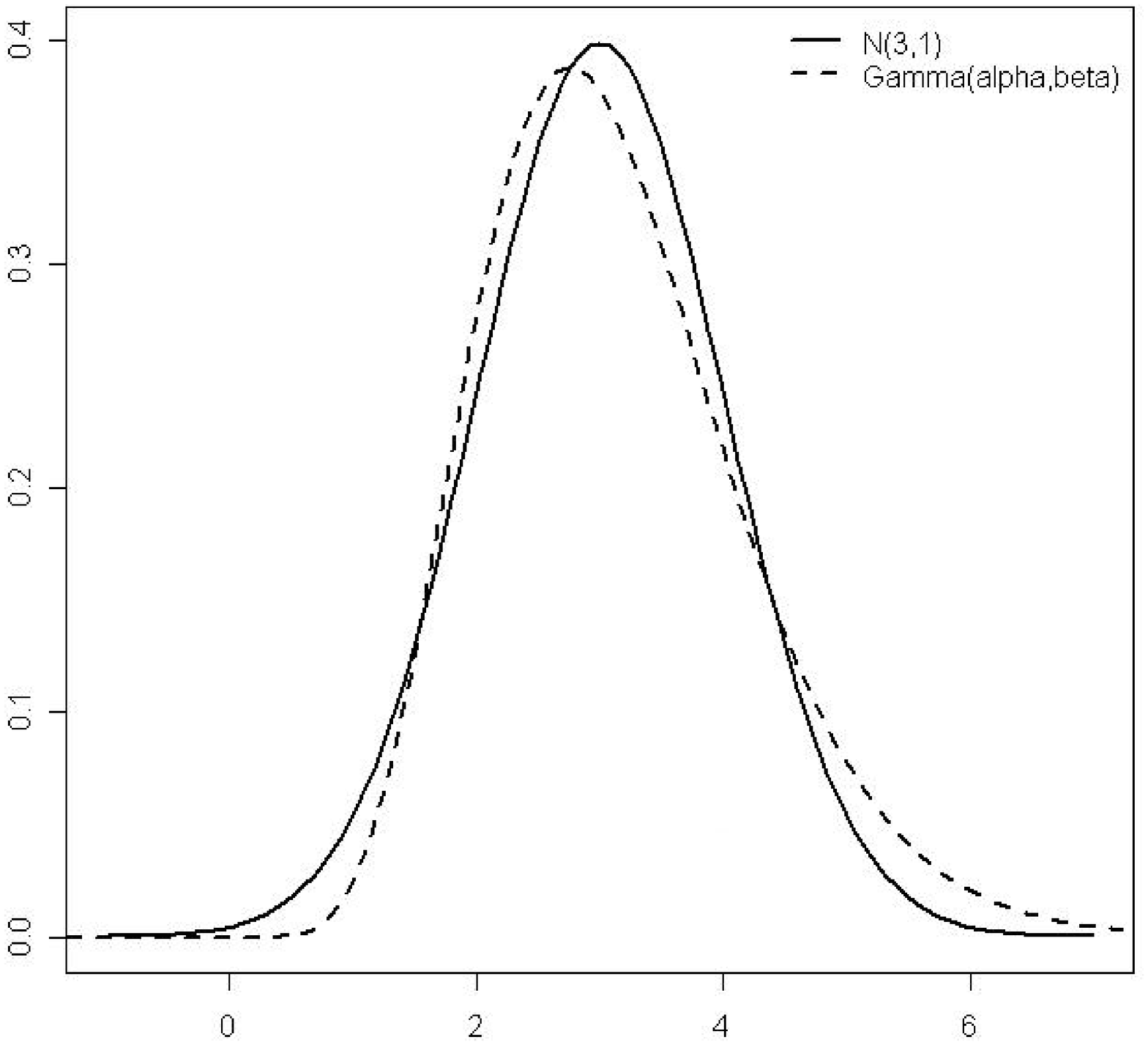}}\hfill
\subfigure[$Wei(2,1)$]{\includegraphics[width=0.33\textwidth]{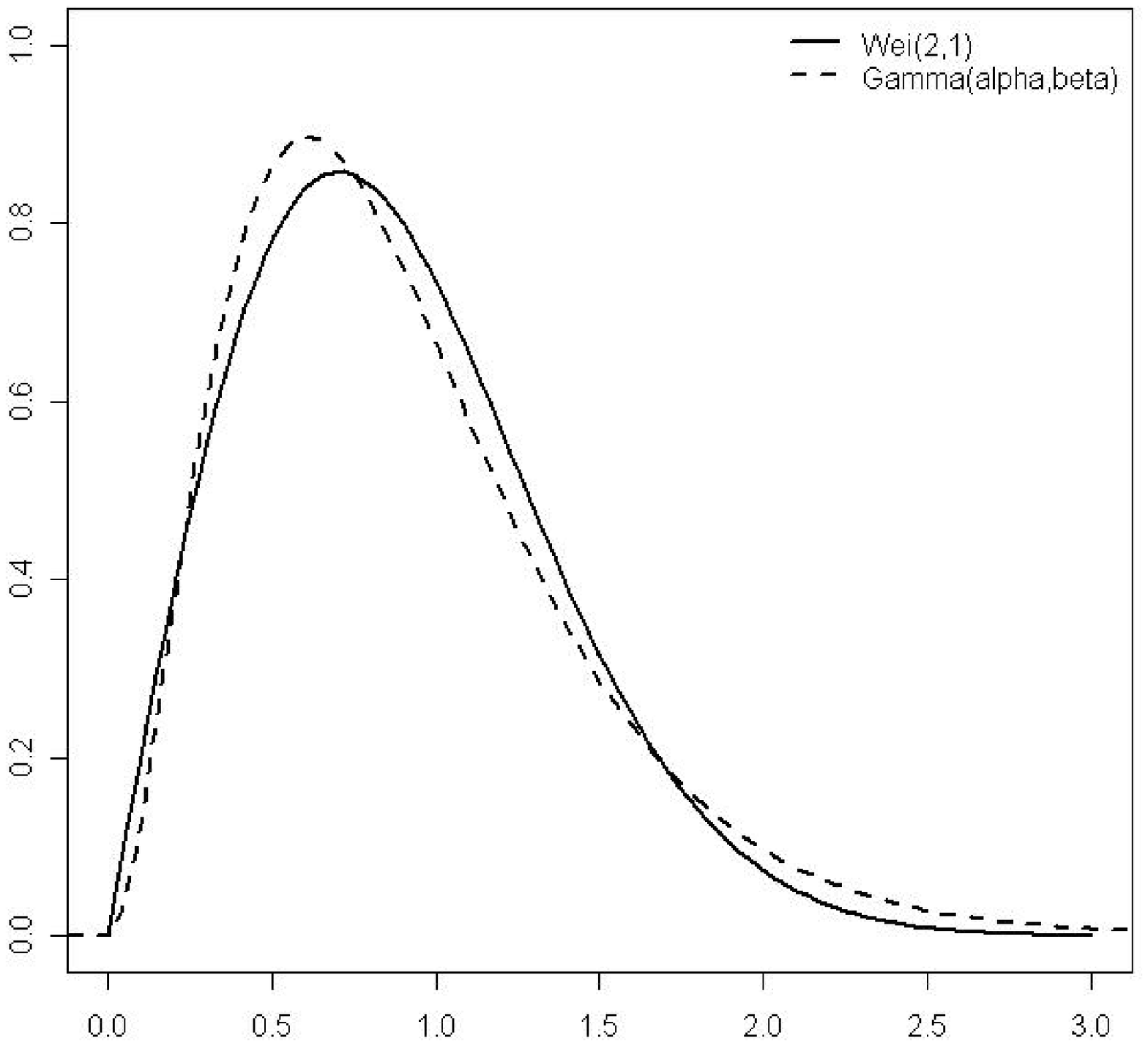}}\hfill
\subfigure[$IG(4,1)$ ]{\includegraphics[width=0.33\textwidth]{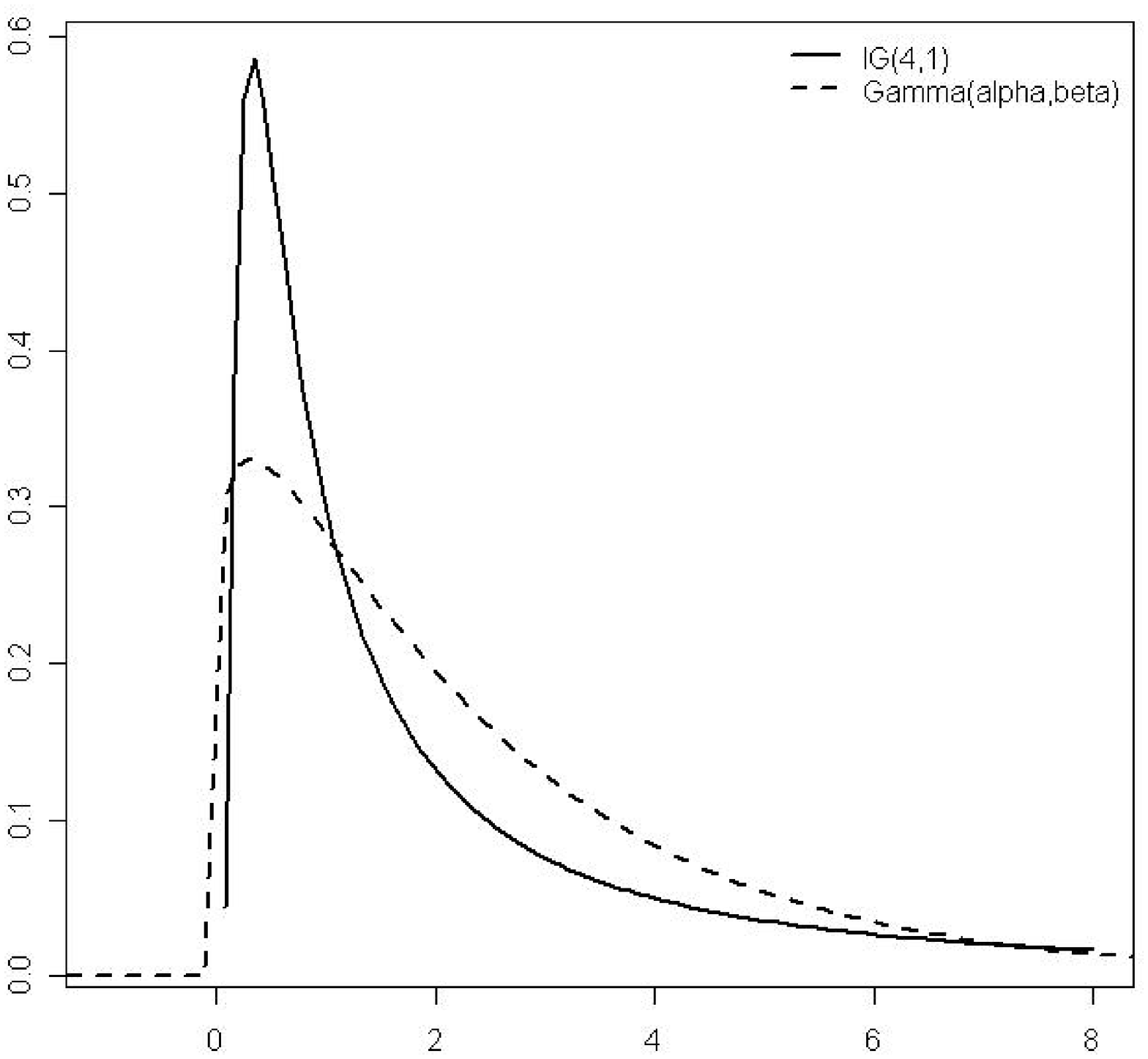}}\hfill
\subfigure[$IG(1,4)$ ]{\includegraphics[width=0.33\textwidth]{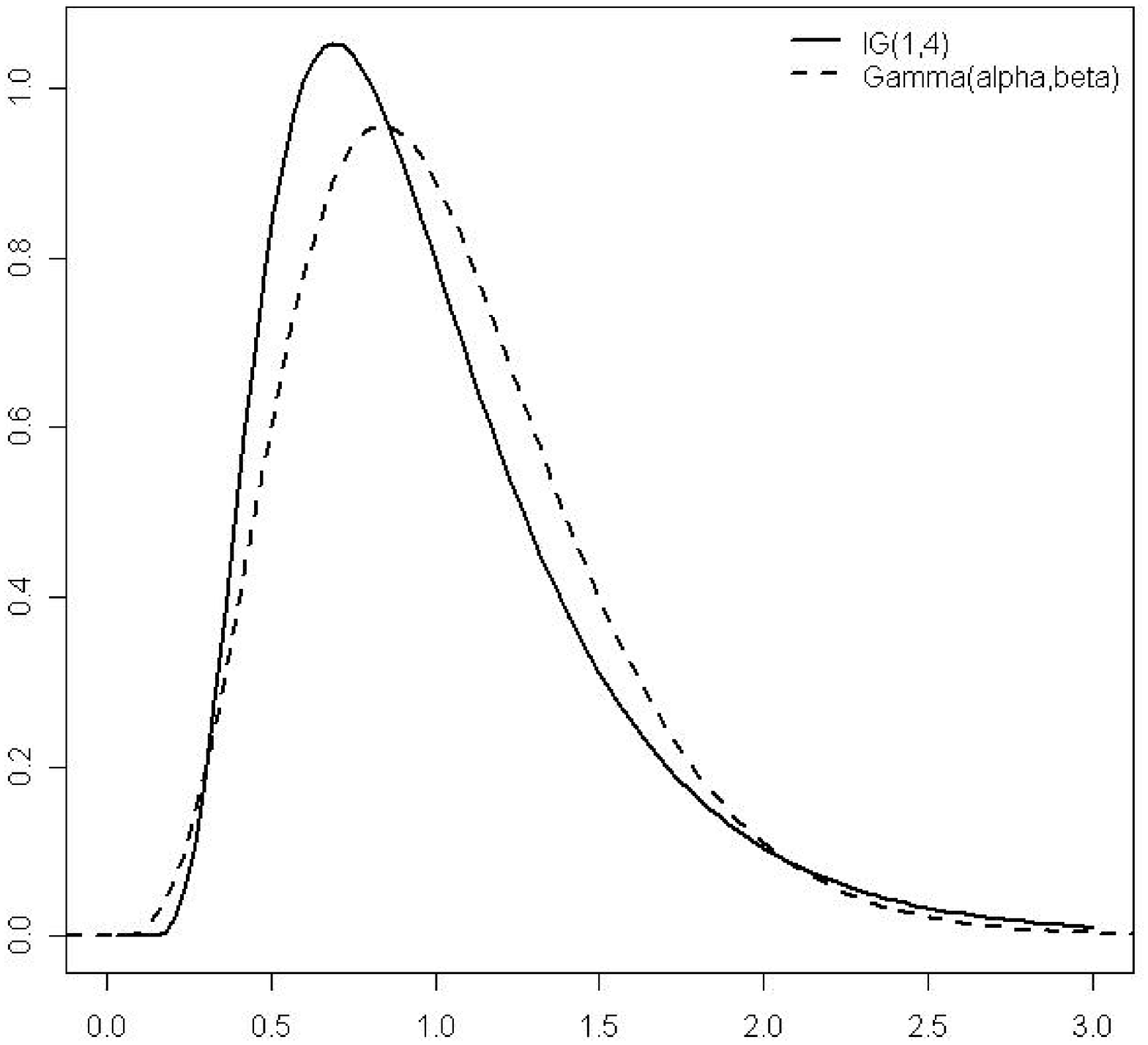}}\hfill
\caption{Densities of some alternative distributions compared with the density of a fitted gamma distribution.
Parameter estimation is based on a sample from the corresponding alternative with $n=100$ and $r=75$.}
\label{Vergleich mit Gammaverteilung}
\end{figure}

\begin{description}
\item[Level:] The tests based on the OS and LHB transformations maintain the nominal level very well, while those based on the MS, FK1 and FK2 are somewhat conservative.

\item[\boldmath $\mathcal{L}\gamma(\alpha,\beta)$ alternatives:] (Figure \ref{Vergleich mit Gammaverteilung}(a),(b))
All variants of MS and OS based tests have good power against the logarithmic gamma distribution, while the FK1--based tests are worst in this case.

\item[\boldmath $\mathcal{N}(3,1)$ alternative:] (Figure \ref{Vergleich mit Gammaverteilung}(c))
The tests based on the MS and the OS transformation (in this order) have higher powers against the $\mathcal{N}(3,1)$ distribution.

\item[\boldmath $Wei(\alpha,\beta)$ alternatives:] (Figure \ref{Vergleich mit Gammaverteilung}(d))
Given the fact that Weibull distributions can be approximated quite well by suitable gamma distributions it is not suprising to see that power is low, uniformly over all tests and transformations. Clearly however the OS--based tests stand out as best.

\item[\boldmath $IG(\mu,\lambda)$ and $\mathcal{LN}(\mu,\sigma)$ alternatives:] (Figure \ref{Vergleich mit Gammaverteilung}(e),(f))
The $\mathcal{LN}(0,1)$ and the $IG(1,4)$ distributions can be well fitted by a gamma distribution.  Hence, power against these alternatives is generally low. Against the $IG(4,1)$ alternative the tests based on MS and OS work best.

\item[Summary:]
Several of the alternatives are difficult to distinguish from a fitted gamma distribution. The best results are observed with the tests based on the transformations of Michael and Schucany and O'Reilly and Stephens, and within these transformations the test based on the characteristic function have a certain edge.
\end{description}

\renewcommand{\baselinestretch}{1.0}
\setlength{\tabcolsep}{1mm}
\begin{landscape}
\begin{table}
\begin{center}

\begin{tabular}{ccc|ccc|ccc|ccc|ccc|ccc}
  \hline
 \multirow{2}{*}{Distribution} & \multirow{2}{*}{$\frac{r}{n}*100\%$} & \multirow{2}{*}{$n$} & \multicolumn{3}{c|}{MS} & \multicolumn{3}{c|}{OS} & \multicolumn{3}{c|}{LHB} & \multicolumn{3}{c|}{FK1} & \multicolumn{3}{c}{FK2} \\
 & & & $A^2$ & $W^2$ & \multicolumn{1}{c|}{$C^2$} & $A^2$ & $W^2$ & \multicolumn{1}{c|}{$C^2$}
     & $A^2$ & $W^2$ & \multicolumn{1}{c|}{$C^2$} & $A^2$ & $W^2$ & \multicolumn{1}{c|}{$C^2$}
     & $A^2$ & $W^2$ & $C^2$ \\
  \hline\hline
	\multirow{4}{*}{$\gamma(2,1)$}
  & 50 & 40 & 4 & 4 & 3 & 6 & 6 & 5 & 5 & 5 & 4 & 3 & 4 & 3 & 3 & 4 & 3 \\
  & 50 &100 & 4 & 4 & 4 & 6 & 6 & 6 & 5 & 5 & 5 & 4 & 4 & 4 & 4 & 4 & 4 \\ \cline{2-18}
  & 75 & 40 & 4 & 4 & 4 & 6 & 6 & 5 & 5 & 5 & 5 & 4 & 4 & 3 & 4 & 4 & 4 \\
  & 75 &100 & 5 & 5 & 4 & 6 & 6 & 6 & 5 & 5 & 5 & 4 & 5 & 4 & 4 & 5 & 4 \\ \hline
  \multirow{4}{*}{$\mathcal{N}(3,1)$}
  & 50 & 40 & 16 & 15 & 17 & 25 & 23 & 26 & 10 & 10 & 9  & 9  & 9  & 9  & 9  & 9  & 8   \\
  & 50 &100 & 31 & 28 & 34 & 46 & 42 & 49 & 17 & 17 & 17 & 19 & 18 & 19 & 18 & 17 & 19  \\ \cline{2-18}
  & 75 & 40 & 22 & 20 & 24 & 31 & 28 & 34 & 10 & 10 & 10 & 11 & 10 & 11 & 10 & 10 & 11  \\
  & 75 &100 & 41 & 38 & 46 & 56 & 52 & 61 & 18 & 17 & 18 & 22 & 19 & 21 & 21 & 19 & 22  \\ \hline
  \multirow{4}{*}{$\mathcal{LN}(0,1)$}
  & 50 & 40 & 6  & 6  & 6  & 5 & 5 & 4 & 5 & 5 & 4 & 4 & 4 & 3 & 4 & 5 & 4 \\
  & 50 &100 & 9  & 9  & 10 & 9 & 8 & 9 & 5 & 5 & 5 & 4 & 4 & 4 & 5 & 6 & 6 \\ \cline{2-18}
  & 75 & 40 & 11 & 11 & 12 & 9 & 9 & 9 & 5 & 5 & 5 & 4 & 4 & 4 & 6 & 6 & 7 \\
  & 75 &100 & 22 & 20 & 26 & 24& 22& 27& 5 & 5 & 5 & 5 & 5 & 5 & 10& 10& 12\\ \hline
  \multirow{4}{*}{$IG(4,1)$}
  & 50 & 40 & 15 & 14 & 16 & 12 & 11 & 11 & 4 & 5 & 4 & 4 & 4 & 3 & 6  & 7  & 6  \\
  & 50 &100 & 32 & 29 & 37 & 37 & 32 & 41 & 6 & 6 & 6 & 5 & 5 & 5 & 12 & 12 & 14  \\ \cline{2-18}
  & 75 & 40 & 39 & 35 & 44 & 37 & 33 & 39 & 6 & 6 & 6 & 5 & 4 & 4 & 15 & 14 & 17  \\
  & 75 &100 & 77 & 70 & 81 & 85 & 80 & 88 & 11& 10& 13& 7 & 6 & 7 & 38 & 35 & 47  \\ \hline
  \multirow{4}{*}{$IG(1,4)$}
  & 50 & 40 & 5 & 5 & 4 & 4 & 4 & 3 & 5 & 5 & 5 & 3 & 4 & 3 & 4 & 4 & 3 \\
  & 50 &100 & 6 & 6 & 7 & 6 & 6 & 6 & 6 & 5 & 5 & 4 & 4 & 4 & 5 & 5 & 5 \\ \cline{2-18}
  & 75 & 40 & 7 & 7 & 7 & 6 & 6 & 5 & 5 & 5 & 5 & 4 & 4 & 4 & 5 & 6 & 5 \\
  & 75 &100 & 11& 11& 13& 11& 10& 12& 5 & 5 & 5 & 5 & 4 & 4 & 7 & 7 & 8 \\ \hline
     \hline
\end{tabular}

\end{center}
\caption{Percentage of rejection of tests for gamma distribution based on 10000 replications (part 1)}
\label{Allgemeines_Verfahren_(gamma_n=100) (Teil 1)}
\end{table}
\end{landscape}

\begin{landscape}
\begin{table}
\begin{center}

\begin{tabular}{ccc|ccc|ccc|ccc|ccc|ccc}
  \hline
 \multirow{2}{*}{Distribution} & \multirow{2}{*}{$\frac{r}{n}*100\%$} & \multirow{2}{*}{$n$} & \multicolumn{3}{c|}{MS} & \multicolumn{3}{c|}{OS} & \multicolumn{3}{c|}{LHB} & \multicolumn{3}{c|}{FK1} & \multicolumn{3}{c}{FK2} \\
 & & & $A^2$ & $W^2$ & \multicolumn{1}{c|}{$C^2$} & $A^2$ & $W^2$ & \multicolumn{1}{c|}{$C^2$}
     & $A^2$ & $W^2$ & \multicolumn{1}{c|}{$C^2$} & $A^2$ & $W^2$ & \multicolumn{1}{c|}{$C^2$}
     & $A^2$ & $W^2$ & $C^2$ \\
  \hline\hline	
  \multirow{4}{*}{$Wei(2,1)$}
  & 50 & 40 & 4 & 5 & 4 & 9  & 8  & 9  & 5 & 5 & 4 & 3 & 4 & 3 & 3 & 4 & 3 \\
  & 50 &100 & 5 & 6 & 6 & 13 & 11 & 14 & 5 & 5 & 5 & 5 & 5 & 4 & 4 & 5 & 4 \\ \cline{2-18}
  & 75& 40  & 6 & 6 & 7 & 12 & 10 & 13 & 5 & 5 & 4 & 4 & 5 & 4 & 4 & 4 & 4 \\
  & 75 &100 & 9 & 8 & 10& 18 & 16 & 21 & 5 & 5 & 5 & 4 & 5 & 4 & 5 & 6 & 6 \\ \hline
  \multirow{4}{*}{$Wei(4,1)$}
  & 50 & 40 & 7  & 7  & 7  & 14 & 13 & 14 & 5 & 5 & 4 & 4 & 4 & 3 & 4 & 4 & 4 \\
  & 50 &100 & 9  & 8  & 10 & 22 & 19 & 24 & 5 & 5 & 5 & 4 & 5 & 4 & 5 & 5 & 6 \\ \cline{2-18}
  & 75 & 40 & 11 & 10 & 12 & 20 & 18 & 22 & 5 & 5 & 5 & 4 & 5 & 4 & 5 & 5 & 6 \\
  & 75 &100 & 20 & 18 & 23 & 37 & 32 & 43 & 5 & 5 & 5 & 6 & 6 & 5 & 8 & 8 & 10 \\ \hline
  \multirow{4}{*}{$\mathcal{L}\gamma(2,1)$}
  & 50 & 40 & 30 & 27 & 33 & 26 & 23 & 25 & 5  & 6  & 5  & 4 & 4 & 3 & 11 & 11 & 12 \\
  & 50 &100 & 70 & 62 & 73 & 77 & 69 & 79 & 12 & 11 & 13 & 7 & 5 & 6 & 29 & 26 & 36 \\ \cline{2-18}
  & 75 & 40 & 71 & 64 & 73 & 70 & 63 & 71 & 12 & 11 & 13 & 4 & 3 & 4 & 32 & 29 & 37 \\
  & 75 &100 & 98 & 95 & 98 & 99 & 99 & 99 & 36 & 31 & 42 & 8 & 6 & 8 & 77 & 69 & 84 \\ \hline
  \multirow{4}{*}{$\mathcal{L}\gamma(4,1)$}
  & 50 & 40 & 22 & 19 & 24 & 18 & 16 & 18 & 5 & 5 & 4 & 4 & 3 & 3 & 7  & 7  & 8  \\
  & 50 &100 & 51 & 45 & 57 & 58 & 52 & 63 & 8 & 7 & 8 & 6 & 5 & 5 & 17 & 17 & 22 \\ \cline{2-18}
  & 75 & 40 & 58 & 52 & 62 & 57 & 51 & 60 & 9 & 9 & 10& 4 & 3 & 4 & 20 & 19 & 24 \\
  & 75 &100 & 94 & 89 & 95 & 98 & 96 & 98 & 25& 22& 29& 6 & 5 & 6 & 54 & 47 & 63 \\ \hline
   \hline
\end{tabular}

\end{center}
\caption{Percentage of rejection of tests for gamma distribution based on 10000 replications (part 2)}
\label{Allgemeines_Verfahren_(gamma_n=100) (Teil 2)}
\end{table}
\end{landscape}

\renewcommand{\baselinestretch}{1.2}


\subsection{Testing for normality}

Simulation results for testing the hypothesis of normality using the estimates suggested by Gupta (1952) are given in Tables
\ref{Allgemeines_Verfahren_(norm_n=100) (Teil 1)} and \ref{Allgemeines_Verfahren_(norm_n=100) (Teil 2)}.
The conclusions drawn from these results are as follows:

\begin{figure}[htp]
\subfigure[$IG(4,1)$]{\includegraphics[width=0.33\textwidth]{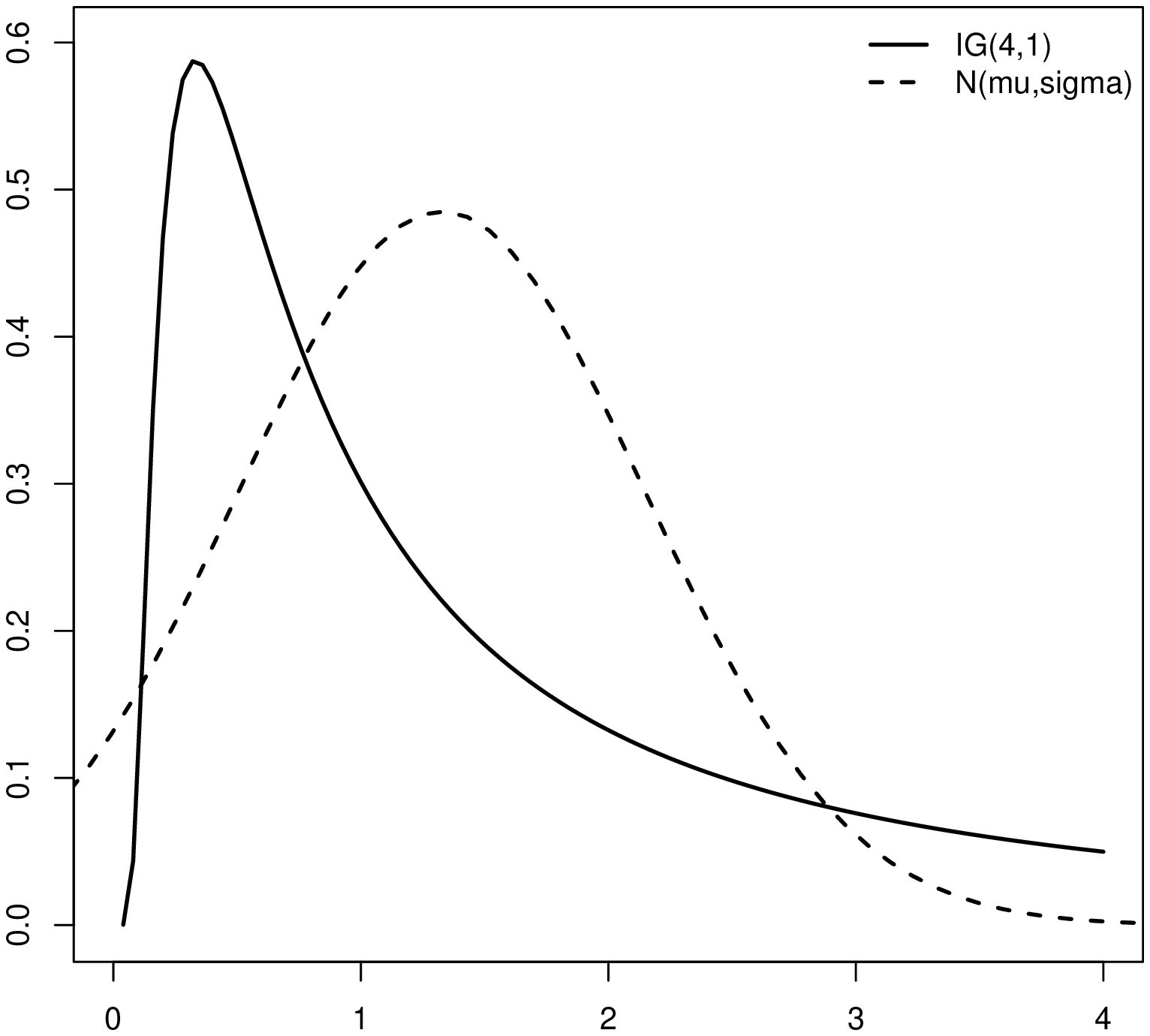}}\hfill
\subfigure[$\gamma(4,1)$ ]{\includegraphics[width=0.33\textwidth]{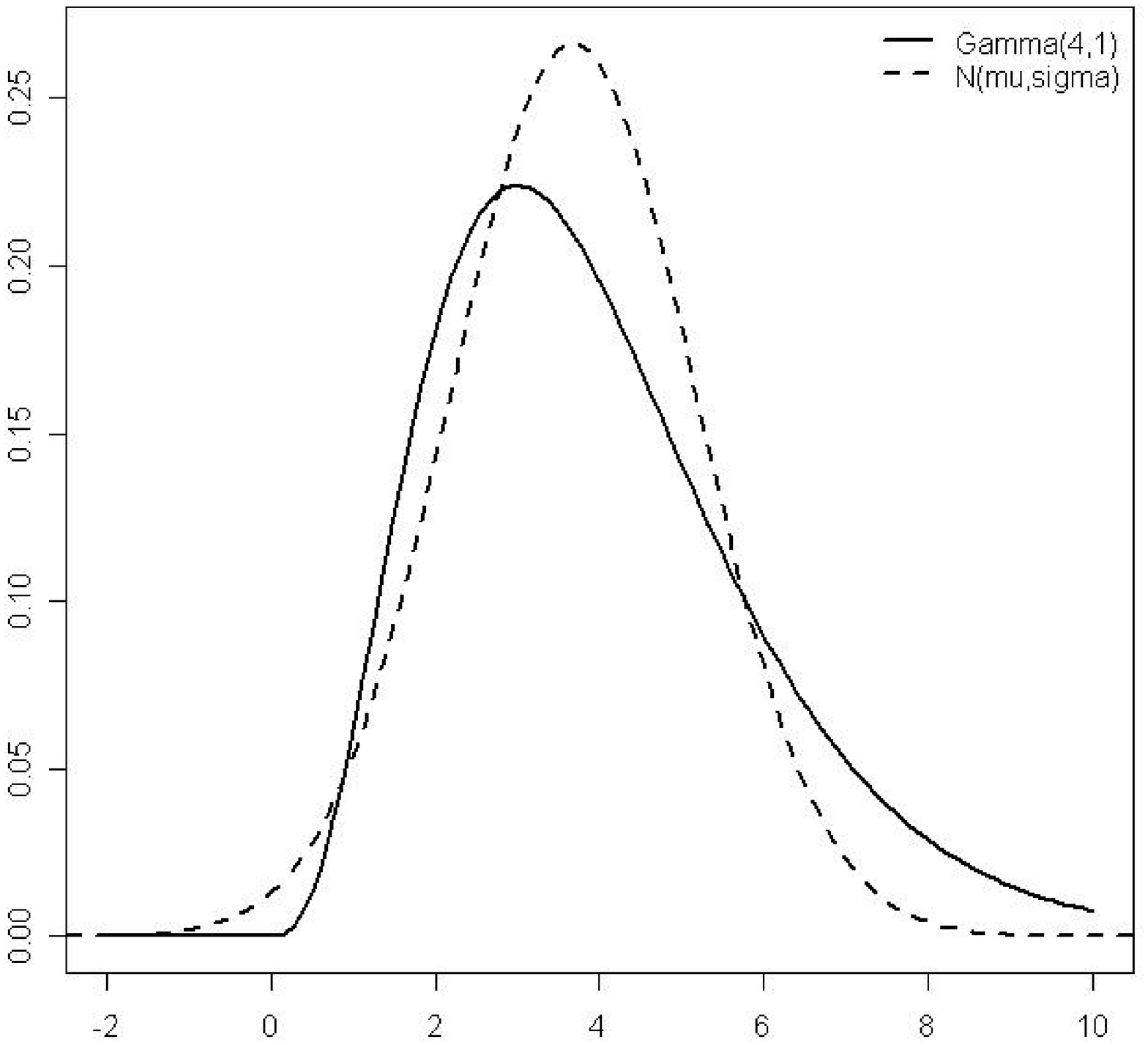}}\hfill
\subfigure[$Wei(2,1)$ ]{\includegraphics[width=0.33\textwidth]{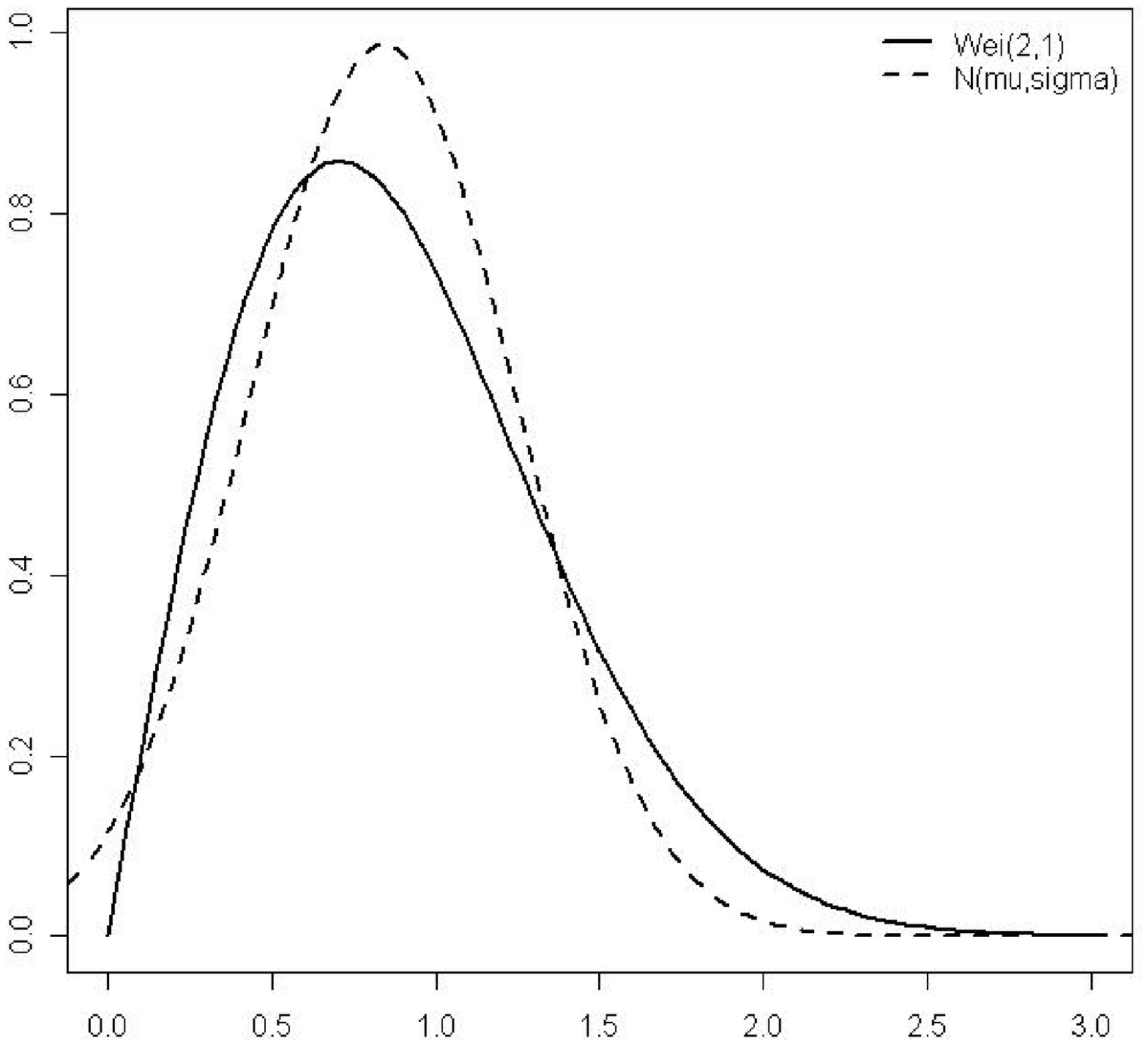}}\hfill
\caption{Densities of some alternative distributions compared with the density of a fitted normal distribution.
Parameter estimation is based on a sample from the corresponding alternative with $n=100$ and $r=75$.}
\label{Vergleich mit Normalverteilung}
\end{figure}

\begin{description}
\item[Level]: Apart from the tests based on FK1, which are somewhat conservative, all tests maintain the nominal level very well.

\item[\boldmath $IG(\mu,\lambda)$ and $\mathcal{LN}(\mu,\sigma)$ alternatives:] (Figure \ref{Vergleich mit Normalverteilung}(a))
For these alternatives, the MS, OS and FK2--based tests have higher powers than LIN and FK1-based tests.

\item[\boldmath $t_m$ alternatives:]
The tests based on MS and OS transforms give the best results, with powers being significantly lower for the $t_4$ alternative.

\item[\boldmath $Exp(\lambda)$ alternative:]
The tests based on the MS, OS and FK2 transformations detect this alternative reliably for medium and low censoring ($r\ge n/2)$ and are clearly preferable to tests based on LHB and FK1.

\item[\boldmath $\gamma(\alpha,\beta)$ and $\mathcal{L}\gamma(\alpha,\beta)$ alternatives:] (Figure \ref{Vergleich mit Normalverteilung}(b))
The tests based on MS, OS and FK2 transformation have high power against logarithmic gamma distributions for medium and low censoring. They are also preferable against gamma alternatives.

\item[\boldmath $Wei(\alpha,\beta)$ alternatives:] (Figure \ref{Vergleich mit Normalverteilung}(c))
Due to the similarity of the $Wei(3,1) $ and $Wei(3,5)$ and suitable normal densities, power is only slightly above the nominal level.

\item[\boldmath $\mathcal{L}(\alpha,\beta)$ alternatives:]
Again, this alternative is hard to distinguish from a normal distribution. Nevertheless, MS and OS based tests show some power.

\item[Summary:]
The highest power is observed with the tests based on the transformations of Michael and Schucany and O'Reilly and Stephens. However, the tests based on transformation FK2 also show a comparable behavior.

As in the case of the exponential distribution, we added results for the direct statistics (DS), the Cram\'er-von Mises and the Anderson-Darling test;
see Section 4.8.4 of D'Agostino and Stephens (1986) for the normal distribution with censored data.
Corresponding results are given in the last two columns in Tables \ref{Allgemeines_Verfahren_(norm_n=100) (Teil 1)} and \ref{Allgemeines_Verfahren_(norm_n=100) (Teil 2)}.
Critical values for our censoring proportions 25\% and 50\% and sample sizes $n=40$ and $100$ are provided in Table \ref{Critical_values_normal_censored_Data}.
All comments given at the end of section 5.1 for the exponential case also apply here, although the transformed--based tests are more competitive in this case.
\end{description}

\renewcommand{\baselinestretch}{1.0}
\setlength{\tabcolsep}{2mm}
\begin{table}[htp]
\begin{center}

\begin{tabular}{cc||c|c|c|c|c|c}
  \hline
 \multirow{2}{*}{$n$} &\multirow{2}{*}{$\frac{r}{n}$} & \multicolumn{3}{c|}{$A^2_{r,n}$} & \multicolumn{3}{c}{$W^2_{r,n}$}\\
 & & $10\%$ & $5\%$ & \multicolumn{1}{c|}{$1\%$} & $10\%$ & $5\%$ & \multicolumn{1}{c}{$1\%$}\\
  \hline\hline	
 40 & 0.50 &  0.222 & 0.286  & 0.493  & 0.036  & 0.049  & 0.097  \\
 40 & 0.75 &  0.364 &  0.451 &  0.683 &  0.064 &  0.080 & 0.120  \\ \hline
100 & 0.50 &  0.227 & 0.285  & 0.446 & 0.036 & 0.047 & 0.083   \\
100 & 0.75 & 0.369 & 0.453 & 0.662 & 0.066 & 0.081 & 0.120   \\ \hline
\end{tabular}

\end{center}
\caption{Critical values of direct edf statistics for testing for normality based on $10^7$ replications}
\label{Critical_values_normal_censored_Data}
\end{table}	

\renewcommand{\baselinestretch}{1.2}

We repeated the simulations using modified maximum likelihood estimation as suggested by Tiku (1967).
The results have been quite similar but power was slightly worse compared to the method of Gupta.
Therefore, the results have been omitted.

\renewcommand{\baselinestretch}{1.0}
\setlength{\tabcolsep}{1mm}
\begin{landscape}
\begin{table}
\begin{center}

\begin{tabular}{ccc|ccc|ccc|ccc|ccc|ccc|cc}
  \hline
 \multirow{2}{*}{Distribution} & \multirow{2}{*}{$\frac{r}{n}*100\%$} & \multirow{2}{*}{$n$} & \multicolumn{3}{c|}{MS} & \multicolumn{3}{c|}{OS} & \multicolumn{3}{c|}{LHB} & \multicolumn{3}{c|}{FK1} & \multicolumn{3}{c|}{FK2} & \multicolumn{2}{c}{DS} \\
 & & & $A^2$ & $W^2$ & \multicolumn{1}{c|}{$C^2$} & $A^2$ & $W^2$ & \multicolumn{1}{c|}{$C^2$}
     & $A^2$ & $W^2$ & \multicolumn{1}{c|}{$C^2$} & $A^2$ & $W^2$ & \multicolumn{1}{c|}{$C^2$}
     & $A^2$ & $W^2$ & \multicolumn{1}{c|}{$C^2$} &$A^2_{r,n}$ & $W^2_{r,n}$ \\
  \hline\hline
	\multirow{4}{*}{$\mathcal{N}(0,1)$}
  & 50 & 40 & 6 & 5 & 5 & 6 & 5 & 5 & 5 & 5 & 5 & 3 & 3 & 3 & 5 & 5 & 5 & 5 & 5\\
  & 50 &100 & 5 & 5 & 5 & 6 & 6 & 6 & 5 & 5 & 5 & 4 & 4 & 3 & 5 & 5 & 5 & 5 & 5\\ \cline{2-20}
  & 75 & 40 & 6 & 5 & 6 & 6 & 6 & 6 & 5 & 5 & 5 & 4 & 4 & 3 & 5 & 5 & 5 & 5 & 5\\
  & 75 &100 & 5 & 5 & 5 & 6 & 5 & 5 & 5 & 5 & 4 & 4 & 4 & 3 & 5 & 5 & 5 & 5 & 5\\ \hline
  \multirow{4}{*}{$\mathcal{LN}(0,1)$}
  & 50 & 40 & 50 & 44 & 53 & 33 & 30 & 32 & 9  & 8  & 10 & 9  & 8  & 8  & 42 & 39 & 41 & 65 & 66 \\
  & 50 &100 & 87 & 78 & 87 & 85 & 78 & 86 & 19 & 18 & 21 & 15 & 12 & 13 & 81 & 76 & 83 & 98 & 98 \\ \cline{2-20}
  & 75 & 40 & 85 & 79 & 85 & 79 & 73 & 79 & 20 & 19 & 22 & 12 & 9  & 11 & 76 & 72 & 78 & 94 & 92 \\
  & 75 &100 & 100& 99 & 100& 100& 100& 100& 55 & 52 & 61 & 21 & 15 & 17 & 99 & 99 & 99 &100 &100 \\ \hline
  \multirow{4}{*}{$IG(4,1)$}
  & 50 & 40 & 75 & 70 & 77 & 60 & 55 & 58 & 16 & 14 & 17 & 12 & 9  & 11 & 65 & 61 & 64 & 88&88\\
  & 50 &100 & 99 & 97 & 99 & 99 & 97 & 99 & 45 & 43 & 49 & 21 & 15 & 17 & 98 & 96 & 98 &100&100 \\ \cline{2-20}
  & 75 & 40 & 99 & 98 & 98 & 98 & 96 & 97 & 49 & 47 & 52 & 20 & 14 & 18 & 96 & 95 & 97 &100&100\\
  & 75 &100 & 100& 100& 100& 100& 100& 100& 93 & 92 & 95 & 29 & 20 & 23 & 100& 100& 100&100&100 \\ \hline
  \multirow{4}{*}{$IG(1,4)$}
  & 50 & 40 & 18 & 16 & 19 & 9  & 9  & 8  & 6 & 6 & 6 & 6 & 5 & 5 & 16 & 15 & 16 & 28 & 30 \\
  & 50 &100 & 33 & 28 & 37 & 28 & 25 & 31 & 7 & 6 & 7 & 9 & 8 & 7 & 32 & 28 & 33 & 59 & 62\\ \cline{2-20}
  & 75 & 40 & 34 & 30 & 38 & 25 & 23 & 27 & 6 & 6 & 7 & 8 & 6 & 7 & 29 & 26 & 30 & 49 & 46 \\
  & 75 &100 & 66 & 56 & 69 & 69 & 62 & 74 & 10& 10& 12& 12& 10& 10& 59 & 53 & 64 & 89 & 87 \\ \hline
  \multirow{4}{*}{$t_2$}
  & 50 & 40 & 50 & 47 & 52 & 58 & 55 & 60 & 10 & 10 & 11 & 2  & 2  & 1  & 5  & 4 & 7  & 37 & 27 \\
  & 50 &100 & 82 & 78 & 84 & 88 & 86 & 90 & 24 & 22 & 26 & 25 & 15 & 24 & 14 & 8 & 20 & 79 & 74\\ \cline{2-20}
  & 75 & 40 & 56 & 53 & 59 & 61 & 58 & 64 & 13 & 12 & 14 & 17 & 11 & 16 & 7  & 5 & 9  & 53 & 51 \\
  & 75 &100 & 87 & 85 & 89 & 90 & 88 & 91 & 33 & 29 & 35 & 47 & 33 & 45 & 19 & 11& 21 & 88 & 87 \\ \hline
  \multirow{4}{*}{$t_4$}
  & 50 & 40 & 20 & 18 & 22 & 26 & 23 & 28 & 6 & 6 & 5 & 3 & 3 & 2 & 4 & 4 & 4 & 13 & 8 \\
  & 50 &100 & 37 & 32 & 40 & 46 & 41 & 50 & 7 & 6 & 7 & 6 & 5 & 6 & 6 & 5 & 9 & 35 & 28 \\ \cline{2-20}
  & 75 & 40 & 23 & 20 & 25 & 27 & 25 & 30 & 5 & 5 & 5 & 5 & 4 & 5 & 4 & 4 & 5 & 20 & 18\\
  & 75 &100 & 41 & 36 & 44 & 47 & 42 & 51 & 9 & 8 & 9 & 11& 7 & 11& 8 & 6 & 9 & 44 & 41\\ \hline
 	     \hline
\end{tabular}

\end{center}
\caption{Percentage of rejection of tests for normality based on 10000 replications (part 1)}
\label{Allgemeines_Verfahren_(norm_n=100) (Teil 1)}
\end{table}
\end{landscape}

\begin{landscape}
\begin{table}
\begin{center}

\begin{tabular}{ccc|ccc|ccc|ccc|ccc|ccc|cc}
  \hline
 \multirow{2}{*}{Distribution} & \multirow{2}{*}{$\frac{r}{n}*100\%$} & \multirow{2}{*}{$n$} & \multicolumn{3}{c|}{MS} & \multicolumn{3}{c|}{OS} & \multicolumn{3}{c|}{LHB} & \multicolumn{3}{c|}{FK1} & \multicolumn{3}{c|}{FK2} & \multicolumn{2}{c}{DS} \\
 & & & $A^2$ & $W^2$ & \multicolumn{1}{c|}{$C^2$} & $A^2$ & $W^2$ & \multicolumn{1}{c|}{$C^2$}
     & $A^2$ & $W^2$ & \multicolumn{1}{c|}{$C^2$} & $A^2$ & $W^2$ & \multicolumn{1}{c|}{$C^2$}
     & $A^2$ & $W^2$ & \multicolumn{1}{c|}{$C^2$} &$A^2_{r,n}$ & $W^2_{r,n}$ \\
  \hline\hline
  \multirow{4}{*}{$\mathcal{L}(0,1)$}
  & 50 & 40 & 9  & 9  & 10 & 13 & 12 & 14 & 5 & 5 & 5 & 3 & 3 & 2 & 4 & 4 & 4 & 5 & 4 \\
  & 50 &100 & 13 & 11 & 15 & 20 & 17 & 22 & 5 & 5 & 5 & 4 & 4 & 3 & 5 & 5 & 6 & 11& 8\\ \cline{2-20}
  & 75 & 40 & 11 & 10 & 11 & 13 & 11 & 15 & 5 & 5 & 5 & 4 & 4 & 3 & 4 & 4 & 5 & 8 & 7\\
  & 75 &100 & 15 & 13 & 17 & 19 & 16 & 22 & 5 & 5 & 5 & 4 & 4 & 4 & 5 & 5 & 6 & 16& 14\\ \hline
  \multirow{4}{*}{$Wei(2,1)$}
  & 50 & 40 & 14 & 12 & 14 & 8  & 7  & 6  & 5 & 5 & 5 & 5 & 5 & 4 & 13 & 12 & 12 & 20 & 22 \\
  & 50 &100 & 24 & 20 & 27 & 20 & 17 & 22 & 6 & 6 & 6 & 8 & 7 & 7 & 24 & 21 & 26 & 46 & 48\\  \cline{2-20}
  & 75 & 40 & 17 & 14 & 18 & 12 & 11 & 11 & 5 & 5 & 5 & 6 & 6 & 5 & 15 & 14 & 16 & 27 & 24\\
  & 75 &100 & 33 & 26 & 36 & 34 & 29 & 38 & 6 & 6 & 6 & 8 & 8 & 8 & 29 & 26 & 33 & 59 & 53\\ \hline
  \multirow{4}{*}{$Exp(1)$}
  & 50 & 40 & 61 & 53 & 60 & 44 & 39 & 40 & 10 & 9  & 10 & 10 & 8 & 8  & 57 & 53 & 57 & 78 & 77 \\
  & 50 &100 & 95 & 90 & 94 & 95 & 90 & 94 & 26 & 25 & 30 & 22 & 19& 19 & 95 & 92 & 96 & 100& 100\\ \cline{2-20}
  & 75 & 40 & 83 & 75 & 81 & 77 & 70 & 74 & 16 & 15 & 17 & 12 & 9 & 10 & 78 & 74 & 80 & 94 & 91\\
  & 75 &100 & 100& 99 & 99 & 100& 99 & 100& 49 & 45 & 55 & 27 & 22& 25 & 100& 99 & 100& 100& 100 \\ \hline
  \multirow{4}{*}{$\gamma(2,1)$}
  & 50 & 40 & 27 & 23 & 29 & 15 & 14 & 13 & 6 & 6 & 6 & 7 & 6 & 5 & 23 & 22 & 23 & 40 & 42 \\
  & 50 &100 & 55 & 45 & 58 & 51 & 43 & 54 & 9 & 8 & 10& 12& 11& 10& 52 & 46 & 55 & 82 & 83 \\ \cline{2-20}
  & 75 & 40 & 45 & 38 & 47 & 35 & 30 & 36 & 7 & 7 & 7 & 9 & 7 & 7 & 37 & 35 & 40 & 62 & 57\\
  & 75 &100 & 82 & 71 & 83 & 86 & 78 & 88 & 15& 14& 17& 14& 12& 12& 76 & 70 & 80 & 97 & 95 \\ \hline
  \multirow{4}{*}{$\gamma(4,1)$}
  & 50& 40 & 14 & 12 & 15 & 7  & 7  & 5  & 6 & 6 & 5 & 6 & 6 & 5 & 13 & 13 & 13 & 20 & 22 \\
  & 50&100 & 23 & 19 & 26 & 19 & 16 & 20 & 6 & 6 & 6 & 8 & 7 & 6 & 22 & 20 & 23 & 44 & 47 \\ \cline{2-20}
  & 75& 40 & 22 & 19 & 24 & 15 & 14 & 15 & 6 & 6 & 6 & 7 & 6 & 5 & 18 & 17 & 19 & 32 & 30\\
  & 75&100 & 42 & 35 & 47 & 44 & 38 & 50 & 7 & 6 & 7 & 10& 8 & 9 & 37 & 33 & 41 & 68 & 64 \\ \hline
  \multirow{4}{*}{$\mathcal{L}\gamma(2,1)$}
  & 50 & 40 & 80 & 74 & 81 & 66 & 61 & 64 & 19 & 17 & 21 & 13 & 10 & 12 & 72 & 68 & 71 & 90  & 90 \\
  & 50 &100 & 99 & 98 & 99 & 99 & 98 & 99 & 51 & 49 & 56 & 23 & 17 & 18 & 99 & 98 & 99 & 100 & 100\\ \cline{2-20}
  & 75 & 40 & 99 & 99 & 99 & 99 & 98 & 99 & 60 & 58 & 63 & 23 & 17 & 23 & 98 & 97 & 98 & 100 & 100 \\
  & 75 &100 & 100& 100& 100& 100& 100& 100& 97 & 96 & 98 & 32 & 23 & 28 & 100& 100& 100& 100 & 100\\ \hline
	     \hline
\end{tabular}

\end{center}
\caption{Percentage of rejection of tests for normality based on 10000 replications (part 2)}
\label{Allgemeines_Verfahren_(norm_n=100) (Teil 2)}
\end{table}
\end{landscape} 

\renewcommand{\baselinestretch}{1.2}

\subsection{Summary of simulation results}

In this subsection, we try to convey a qualitative message based on the simulation results.
Table \ref{rank-table} has the three tests as entries for the lines, and the five transformations to uniformity
as entries for the columns. At each cell, i.e. at each specific combination of test and transformation, there are
two entries which show order: The left digit is the order w.r.t. the lines (tests), while the right digit shows
the order w.r.t. the columns (transformations).
For example the entry $2\backslash 1$ in the first cell means that the Anderson-Darling test is the second best for the MS transformation, while the MS transformation ranks 1st for the Anderson-Darling test.

To get such an overall assessment for, say, a fixed transformation, we assigned ranks
to the three tests. The test with the highest percentage of rejection has rank one, and so on.
Then, we summed up the ranks for all listed combinations of null hypotheses, alternatives and censoring proportions $r$.
The test with the lowest sum score has the entry 1 in Table \ref{rank-table} for this transformation.

This overview gives a clear picture: For any given test, the MS and OS transformations perform best. In this connection, a direct look at the sum scores shows that there is not much to choose between these two transforms. On the other hand, the LHB transform follows at a clear distance, having the edge over FK2. By far the lowest sum score for all tests has the FK1 transformation. Also, for the MS and OS transformations, the best test is the characteristic function test $C^2$. $C^2$ also performs best for LHB and FK2 transform. Finally, the best combination of test and transformation is $C^2$ with OS.


\vspace*{10mm}

\renewcommand{\baselinestretch}{0.9}
\setlength{\tabcolsep}{1mm}
\begin{table}[htp]
\centering
\begin{tabular}{c|c|c|c|c|c}
  \hline
  \multirow{3}{*}{\backslashbox[10mm]{Test}{Transformation}}
 & \multirow{3}{*}{MS} & \multirow{3}{*}{OS} &\multirow{3}{*}{LHB} & \multirow{3}{*}{FK1} & \multirow{3}{*}{FK2} \\
 &&&&&\\
 &&&&&\\ \hline
\multirow{3}{*}{$A^2$} & \multirow{3}{*}{\backslashbox[10mm]{2}{\textbf{1}}} & \multirow{3}{*}{\backslashbox[10mm]{2}{2}} &\multirow{3}{*}{\backslashbox[10mm]{2}{3}}&\multirow{3}{*}{\backslashbox[10mm]{\textbf{1}}{5}}&\multirow{3}{*}{\backslashbox[10mm]{2}{4}}\\
&&&&&\\
&&&&&\\
\hline
\multirow{3}{*}{$W^2$} & \multirow{3}{*}{\backslashbox[10mm]{3}{2}} & \multirow{3}{*}{\backslashbox[10mm]{3}{\textbf{1}}} &\multirow{3}{*}{\backslashbox[10mm]{3}{3}}&\multirow{3}{*}{\backslashbox[10mm]{3}{5}}&\multirow{3}{*}{\backslashbox[10mm]{3}{4}}\\
&&&&&\\
&&&&&\\
\hline
\multirow{3}{*}{$C^2$} & \multirow{3}{*}{\backslashbox[10mm]{\textbf{1}}{2}} & \multirow{3}{*}{\backslashbox[10mm]{\textbf{1}}{\textbf{1}}} &\multirow{3}{*}{\backslashbox[10mm]{\textbf{1}}{3}}&\multirow{3}{*}{\backslashbox[10mm]{2}{5}}&\multirow{3}{*}{\backslashbox[10mm]{\textbf{1}}{4}}\\
&&&&&\\
&&&&&\\
\hline
\end{tabular}
\caption{Overall assessment for the tests and transformations. Left digit: order w.r.t. the lines (tests). Right digit: order w.r.t. the columns (transformations).}
\label{rank-table}
\end{table}

\renewcommand{\baselinestretch}{1.2}

\section{Conclusion}\label{sec_6}
We have applied a series of transformations to original type--II censored data with the aim of  rendering  corresponding full--sample tests statistics for the parent population, applicable and approximately distribution--free. The conclusions from our Monte Carlo study show that these transformations generally work well across all goodness--of--fit tests studied in terms of recovering the nominal level of significance. On the other hand, the best transformation in terms of power depends on the distribution under test, with the transformation of Lin et al. (2008) being best for testing exponentiality, and the transformations of Michael and Schucany (1988) and O'Reilly and Stephens (1988) for testing normality and testing for the gamma distribution. Also, and within each transformation, the characteristic function based test seem to yield the best power for the majority of alternatives. However, this superiority is generally not significant compared to the large differences between the transformations.
Before closing however we wish to reiterate once more that, despite the fact that the transformed based tests often show good power, their advantage lies not so much in the power but in the general applicability of the method: The user does not need new critical values for each distribution under test, each combination of sample size and  censoring proportion, and each possible choice of parameter value/parameter estimate, but,  following the transformation suggested, essentially faces the simplified problem of testing for normality with estimated parameters.

\begin{appendix}
\section{Appendix}

In this appendix we shall investigate the reasons underlying the eventual validity of the Chen--Balakrishnan transformation. In doing so, first in Appendix A.1 we report results on the process corresponding to goodness--of--fit testing for the normal distribution with estimated parameters. Then in Appendices A.2 and A.3 we study in detail the process produced by the Chen--Balakrishnan transformation and compare it with the process in A.1 both theoretically and by simulation.

\subsection{The empirical process under normality}

Suppose that $Z_j, \ j=1,...,n$, are iid normal with unknown mean and variance. Then most standard goodness--of--fit tests are merely functionals of the empirical process
\begin{eqnarray*}
\hat{\alpha}_n(t) &=& \frac{1}{\sqrt{n}}
  \sum_j \left[ I\left\{ \Phi\left( \frac{Z_j-\bar{Z}}{s_Z} \right) \leq t\right\} - t \right] \\
  &=& \frac{1}{\sqrt{n}} \sum_j \left[ I\left\{ U_j \leq \Phi\left( \bar{Z} + s_Z \; \Phi^{-1}(t) \right) \right\} - t \right],
\end{eqnarray*}
where $\bar Z$ and $s_Z$ are the sample mean and sample variance of $Z_1,...,Z_n$, and  $U_j=\Phi(Z_j)$ and $t \in [0,1]$. This process has been studied by Durbin (1973) and showed that under regularity conditions, $$ \hat{\alpha}_n \Rightarrow \alpha$$
where $\alpha$ is a centered Gaussian process with covariance function
\begin{eqnarray*}
 C(\alpha(s),\alpha(t)) &=& \min(s,t) - st - \varphi(\Phi^{-1}(s)) \varphi(\Phi^{-1}(s)) \\
  && - \frac{1}{2} \Phi^{-1}(s) \varphi(\Phi^{-1}(s)) \Phi^{-1}(t) \varphi(\Phi^{-1}(t)),
\end{eqnarray*}
where $\Phi^{-1}$ and $\varphi$ are the quantile and density function of the standard normal distribution. (We note that Loynes (1980) extended the analysis from the iid setting to the case of generalized linear models). {\it{ Clearly the process $\alpha_n$ is identical to the process involved in the Chen--Balakrishnan transformation only in the case of testing for normality with estimated parameters}}.

\subsection{The empirical process under non--normality}

\begin{figure}[htp]
\begin{center}
\includegraphics[scale=0.7]{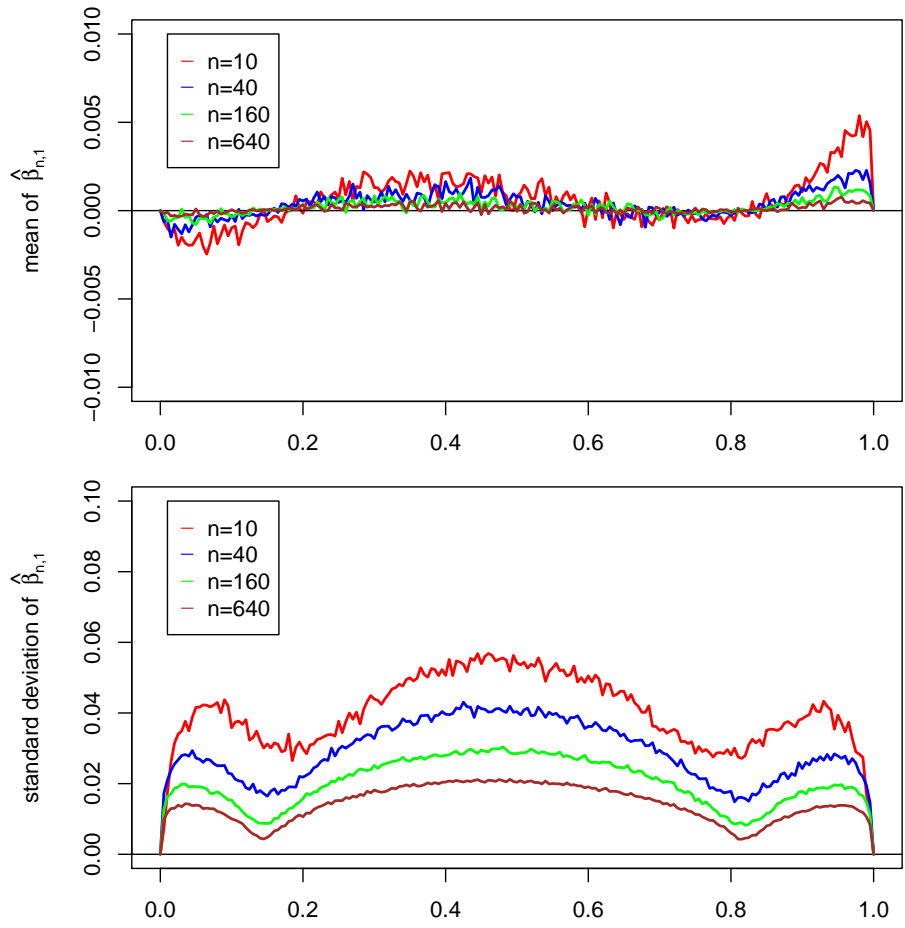}
\caption{\label{figure2} Simulated mean and standard deviation of $\hat{\beta}_{n,1}$ for different sample sizes}
\end{center}
\end{figure}

In this section, we consider iid random variables $X_j$ with DF ${\cal{F}}_{\vartheta}(x)$
(assumed to be continuous and strictly increasing) and the standardized quantile residuals $Z_j= \frac{Y_j-\bar{Y}}{s_Y}$ with
$Y_j= \Phi^{-1} \left( {\cal{F}}_{\hat\vartheta}(X_j) \right)$
(concerning the term {\it standardized quantile residual}, refer to Klar and Meintanis 2011, sec 2.1).
We shall study the following empirical process based on the $Z_j$:
\begin{eqnarray*}
\hat{\beta}_n(t) &=& \frac{1}{\sqrt{n}} \sum_j \left[ I\left\{ \Phi(Z_j) \leq t \right\} - t \right] \\
  &=& \frac{1}{\sqrt{n}}  \sum_j \left[ I\left\{ \Phi\left( \frac{Y_j-\bar{Y}}{s_Y} \right) \leq t\right\} - t \right] \\
  &=& \frac{1}{\sqrt{n}} \sum_j \left[ I\left\{ X_j \leq {\cal{ F}}_{\hat\vartheta}^{-1}
   ( \Phi(\bar{Y} + s_Y \, \Phi^{-1}(t)) ) \right\} - t \right]  \\
  &=& \frac{1}{\sqrt{n}} \sum_j \left[ I\left\{ U_j \leq  {\cal{F}}_{\vartheta} \left(
 {\cal{F}}_{\hat\vartheta}^{-1} \left( \Phi(\bar{Y} + s_Y \, \Phi^{-1}(t)) \right) \right)\right\} - t \right],
\end{eqnarray*}
where ${\cal{ F}}_{\vartheta}^{-1}(p)$ denotes the quantile function of ${\cal{ F}}_{\vartheta}(\cdot)$,
and $U_j = {\cal{F}}_{\vartheta}(X_j)$ are iid uniformly distributed on $[0,1]$. {\it{This is the empirical process actually produced by the Chen--Balakrishnan transformation}}.
Now define $c_Y(t)=\Phi(\bar{Y} + s_Y \Phi^{-1}(t))$, and, similarly,
$c_N(t)=\Phi(\bar{N} + s_N \Phi^{-1}(t))$, where $N_j=\Phi^{-1}(U_j)$ are iid standard normal
random variates, and $\bar{N}$ and $s_N^2$ are the arithmetic mean and sample variance of $N_1,\ldots,N_n$.

\begin{figure}[htp]
\begin{center}
\includegraphics[scale=0.7]{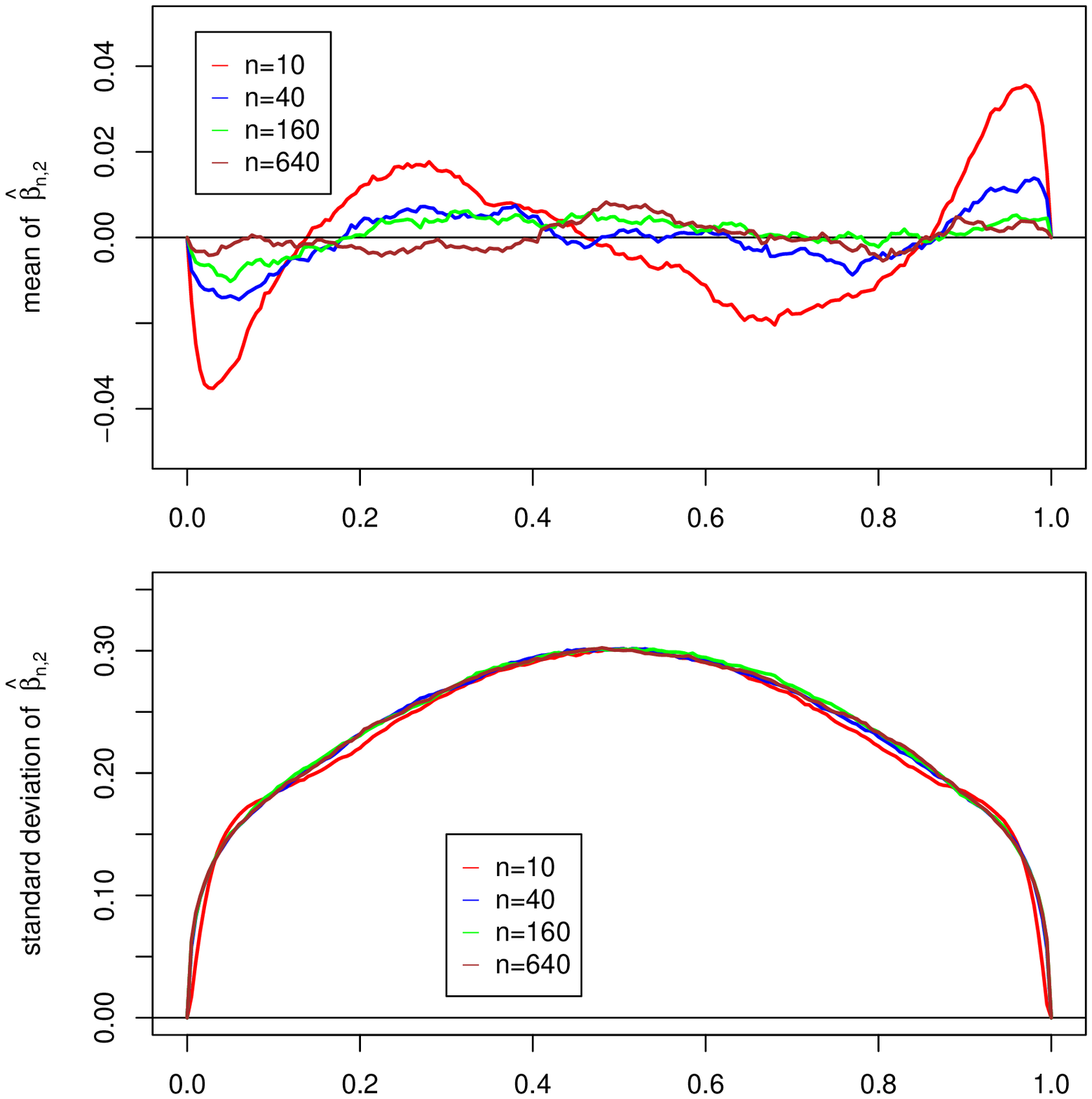}
\caption{\label{figure3} Simulated mean and standard deviation of $\hat{\beta}_{n,2}$ for different sample sizes}
\end{center}
\end{figure}

Then we can decompose the above process as (compare Chen (1991), pp 126-128)
\begin{equation}
\hat{\beta}_n(t) = \hat{\beta}_{n,1}(t) + \hat{\beta}_{n,2}(t) + \hat{\beta}_{n,3}(t),
\end{equation}
where
$$\hat{\beta}_{n,1}(t) = \frac{1}{\sqrt{n}} \sum_j \left[
  I\left\{ U_j \leq  {\cal{F}}_{\vartheta} \left( {\cal{F}}_{\hat\vartheta}^{-1} (c_Y(t)) \right) \right\}
    - {\cal{F}}_{\vartheta} \left( {\cal{F}}_{\hat\vartheta}^{-1} (c_Y(t)) \right)  \right. \\
  - \; I\left\{ U_j \leq c_N(t) \right\} + c_N(t)  \Big],$$
$$\hat{\beta}_{n,2}(t)= \frac{1}{\sqrt{n}} \sum_j \left[  I\left\{ U_j \leq \Phi(\bar{N} + s_N \, \Phi^{-1}(t)) \right\} - t \right],$$ and
$$\hat{\beta}_{n,3}(t)= \frac{1}{\sqrt{n}} \sum_j \left[
   {\cal{F}}_{\vartheta} \left( {\cal{F}}_{\hat\vartheta}^{-1} (c_Y(t)) \right) - c_N(t)  \right].$$

The first part $\hat{\beta}_{n,1}(t)$ in decomposition (A.1) is the difference of an empirical process and a random perturbation thereof, and should be $o_P(1)$
under appropriate regularity conditions (see Chen (1991), p.128, Loynes (1980), Lemma 1, and Rao and Sethuraman (1975)). To check this claim, we simulated a random sample of size $n$ from a unit mean exponential distribution and computed $\hat{\beta}_{n,1}(t), \ t \in [0,1]$, on the basis of an equidistant grid with spacing equal to $0.005$.
This was repeated $B=10000$ times. We approximated the mean function $E[\hat{\beta}_{n,1}(t)]$
and the standard deviation $\sqrt{Var[\hat{\beta}_{n,1}(t)]}$ by the arithmetic mean and
empirical standard deviation based on the $B$ replications; Figure \ref{figure2} shows the result for sample sizes
$n=10,40,160$ and $640$. Clearly the mean function is nearly zero and decreases for increasing $n$, while the standard deviation is small compared to the standard deviation of $\hat{\beta}_n$
or $\hat{\beta}_{n,2}$ (see below). The corresponding variance seems to converge to zero, but rather slowly, with a speed of convergence approximately equal to $1/\sqrt{n}$.

The second part $\hat{\beta}_{n,2}$ corresponds to the normal empirical process $\hat{\alpha}_{n}$ in Appendix A.1.
Figure \ref{figure3} shows the empirical mean function and standard deviation of $\hat{\beta}_{n,2}$ for an underlying
exponential distribution computed in the same way as for $\hat{\beta}_{n,1}$ above.
The mean function, which takes on much larger values than that of $\hat{\beta}_{n,1}$, again converges to zero, whereas the variance function
is nearly constant for $n\ge 40$.

\begin{figure}[htp]
\begin{center}
\includegraphics[scale=0.7]{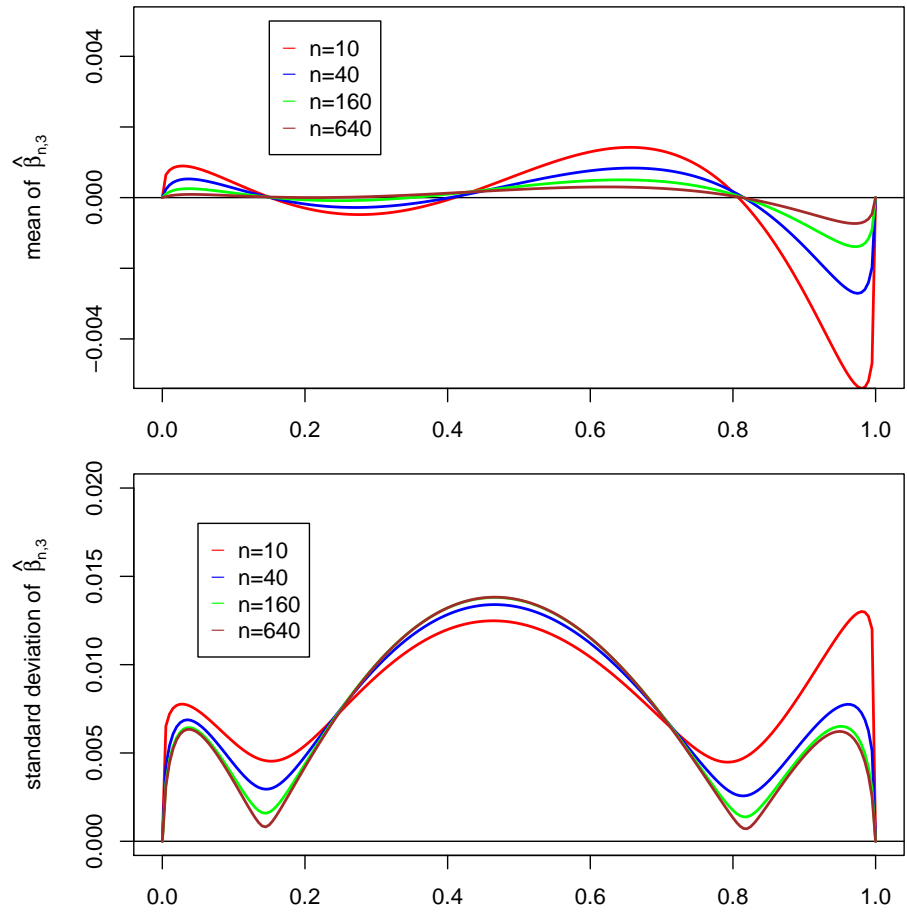}
\caption{\label{figure4} Simulated mean and standard deviation of $\hat{\beta}_{n,3}$ for different sample sizes}
\end{center}
\end{figure}

For the third part we have
$$
 \hat{\beta}_{n,3}(t)= \sqrt{n} \left[ {\cal{F}}_{\vartheta} \left( {\cal{F}}_{\hat\vartheta}^{-1}
  (\Phi(\bar{Y} + s_Y \, \Phi^{-1}(t))) \right) - \Phi(\bar{N} + s_N \, \Phi^{-1}(t)) \right].
$$
In general, this  process does not converge to zero in probability. However, the contribution of $\hat{\beta}_{n,3}$
seems to be negligibly small {\it{in comparison to}} $\hat{\beta}_{n,2}$ in many situations. Figure \ref{figure4} shows the empirical mean function and standard deviation of $\hat{\beta}_{n,3}$ for the exponential distribution, computed as above. The mean function is very small and goes to zero.
The standard deviation is small compared to the standard deviation of $\hat{\beta}_{n,2}$, and it converges, but not to zero. We stress that the crucial point for the behavior of $\hat{\beta}_{n,3}$ is the coupling between the $Y_j$'s and the normal variates $N_j$
which are both based on the original $X_j$, the first computed by using $\hat \vartheta$ while the second by using $\vartheta$. In fact if we generate iid standard normal random variables $\tilde{N}_j$ independent of the $X_j$'s and use them instead of the $N_j$'s, the mean function is small, but does not seem to converge to zero, and the variance is much larger, even larger than that of $\hat{\beta}_{n,2}$.

\begin{figure}[htp]
\begin{center}
\includegraphics[scale=0.7]{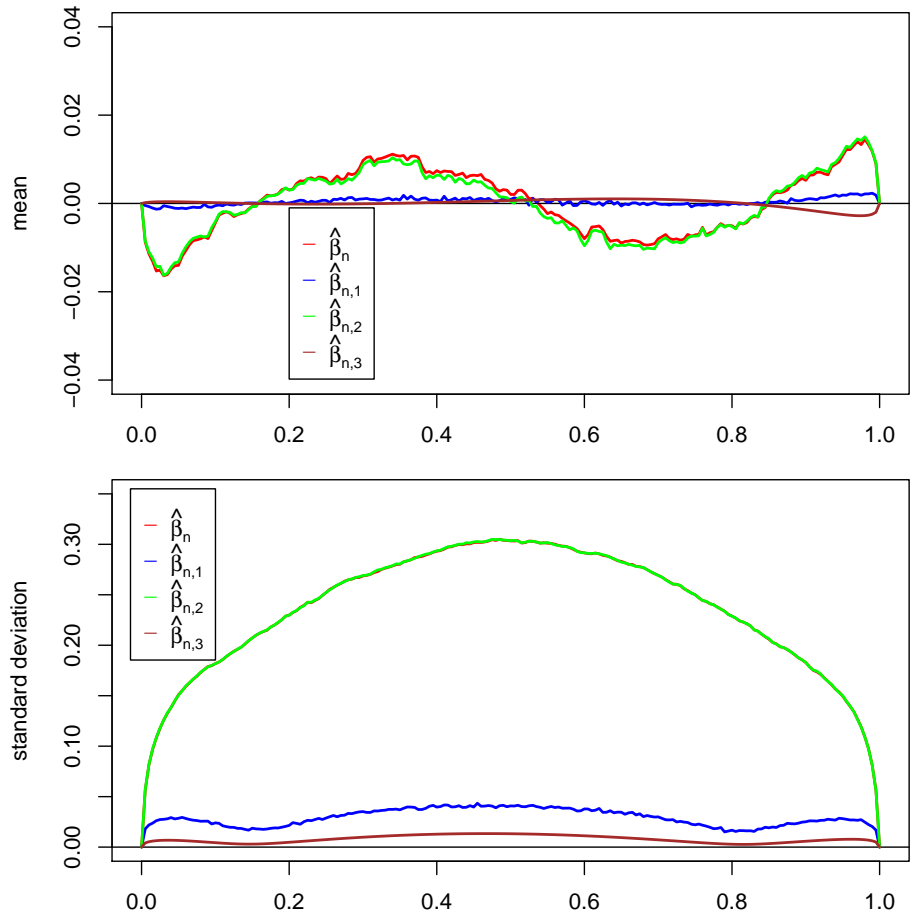}
\caption{\label{figure6} Simulated mean and standard deviation of the different processes for sample size $n=40$}
\end{center}
\end{figure}

From the above it follows that  the values of the process $\hat{\beta}_{n}$ will be eventually dominated by $\hat{\beta}_{n,2}$, at least for large $n$. This is documented in Figure \ref{figure6} where the mean and standard deviation of all four processes are plotted for $n=40$. Note that the
standard deviations of $\hat{\beta}_{n}$ (in red) and $\hat{\beta}_{n,2}$ (in green) are nearly identical, and therefore, visually indistinguishable.

\subsection{Further analysis of the process $\boldsymbol{\hat{\beta}_{n,3}}$}

To keep things simple, we assume in the following that $\vartheta\in \Theta \subset \R$.
Let $\vartheta_0$ denote the true parameter value, and define
\bq
 N_j(\vartheta) &=& \Phi^{-1}\left( {\cal{F}}_{\vartheta}\left( X_j \right) \right), \\
 \bar{N}(\vartheta) &=& \frac1n \sum_{j=1}^n N_j(\vartheta), \quad
 s_N^2(\vartheta) \;=\; \frac{1}{n-1} \sum_{j=1}^n \left(N_j(\vartheta)-\bar{N}(\vartheta)\right)^2.
\eq
Then, $N_j(\vartheta_0)=N_j, \bar{N}(\vartheta_0)=\bar{N}, s_N^2(\vartheta_0)=s_N^2$, and
$N_j(\hat\vartheta)=Y_j, \bar{N}(\hat\vartheta)=\bar{Y}, s_N^2(\hat\vartheta)=s_Y^2$.
Putting
\bq
 h_t(\vartheta) &=& \Phi\left( \bar{N}(\vartheta) + s_N(\vartheta)\cdot \Phi^{-1}(t) \right),
\eq
we obtain $h_t(\vartheta_0)=c_N(t)$ and $h_t(\hat\vartheta)=c_Y(t)$. Thus, we can write
\be \label{hatbeta}
 \hat{\beta}_{n,3}(t) &=& \sqrt{n} \left( g_t(\hat\vartheta) - g_t(\vartheta_0) \right),
\ee
where
\bq
 g_t(\vartheta) &=& {\cal{F}}_{\vartheta_0}\left( {\cal{F}}_{\vartheta}^{-1}\left( h_t(\vartheta) \right) \right).
\eq
Assume now that $\sqrt{n}(\hat\vartheta-\vartheta_0)=O_p(1)$. Then, by using the expansion
\bq
 g_t(\hat\vartheta) = g_t(\vartheta_0) + (\hat\vartheta-\vartheta_0) \, g_t^{'}(\vartheta_0) + (\hat\vartheta-\vartheta_0)^2 \, g_t^{''}(\vartheta^*)/2,
\eq
with $\vartheta^*$ between $\hat\vartheta$ and $\vartheta_0$, and by omitting the quadratic term, we see that $\hat{\beta}_{n,3}(t)$ can be
approximated by
\be \label{ringbeta}
  \mathring{\beta}_{n,3}(t) &=& \sqrt{n} \left(\hat\vartheta-\vartheta_0 \right) \, g_t^{'}\left(\vartheta_0\right).
\ee
Of course, the validity of such a Taylor expansion is not enough to justify the uniform convergence
$\sup_t |\hat{\beta}_{n,3}(t) - \mathring{\beta}_{n,3}(t)|=o_P(1)$.
A sufficient condition would be Fr\'echet differentiability of $g_t(\cdot)$ (see, e.g. van der Vaart and Wellner (1996), p. 373).
However, since we do not intend to give rigorous theory here, this issue is not discussed in any detail. Further analysis of $g_t^{'}\left(\vartheta_0\right)$ leads to the following result, the proof of which is omitted.

\begin{lem}
Let $\bar W$ and $s_W^2$ denote the arithmetic mean and sample variance of the random variables $W_{j0}:=W_j(\vartheta_0)$,
with $W_j(\vartheta):=d N_j(\vartheta)/d\vartheta$, while $r$ denotes the sample correlation coefficient of $W_{10},\ldots,W_{n0}$
and $N_1,\ldots,N_n$. Then,
\bq
 g_t^{\prime}(\vartheta_0) &=& \frac{\partial {\cal{F}}_{\vartheta_0}\left( F_{\vartheta_0}^{-1}(c_N(t)) \right)}{\partial x} \cdot
   \left( \frac{\partial {\cal{F}}_{\vartheta_0}^{-1}(c_N(t))}{\partial p} \cdot h_t^{\prime}(\vartheta_0)
          + \frac{\partial {\cal{F}}_{\vartheta_0}^{-1}(c_N(t))}{\partial \vartheta} \right), \\
 h_t^{\prime}(\vartheta_0) &=& \varphi\left( \bar{N}+s_N \, \Phi^{-1}(t) \right) \cdot \left( \bar{W}+\Phi^{-1}(t) \, r \, s_W \right),
\eq
where $\varphi(\cdot)$ denotes the density of the standard normal distribution.
\end{lem}

Since $N_1,\ldots,N_n$ are iid standard normal variates, $\bar{N} \to 0$ and $s_N\to 1$ almost surely.
Furthermore, $\bar{W} \to \mu_W, s_W \to \sigma_W$, and $r \to \rho$ a.s.,
where $(\mu_W,\sigma_W^2)$ are the mean and variance of $W_{10}$, while $\rho$ denotes the correlation coefficient of $W_{10}$ and $N_1$.
Hence, the following approximation holds for the process in (A.3).

\begin{lem}
The process  $\mathring{\beta}_{n,3}(t)$ can be approximated by the process
\be \label{tildebeta}
\tilde{\beta}_{n,3}(t) &=& \sqrt{n} \left(\hat\vartheta-\vartheta_0 \right) \, \tilde{g}_t^{'}\left(\vartheta_0\right),
\ee
where \bq
 \tilde{g}_t^{\prime}(\vartheta_0) &=& \frac{\partial {\cal{F}}_{\vartheta_0} \left( {\cal{F}}_{\vartheta_0}^{-1}(t) \right)}{\partial x} \cdot
   \left( \frac{\partial {\cal{F}}_{\vartheta_0}^{-1}(t)}{\partial t} \cdot \tilde h_t^{\prime}(\vartheta_0)
          + \frac{\partial {\cal{F}}_{\vartheta_0}^{-1}(t)}{\partial \vartheta} \right),
\eq
\bq
 \tilde h_t^{\prime}(\vartheta_0) &=& \varphi\left(\Phi^{-1}(t)\right) \cdot \left( \mu_W+\Phi^{-1}(t) \ \rho \ \sigma_W\right).
\eq
\end{lem}

\vspace{3mm}
Figure \ref{figure7} shows the simulated mean and standard deviation of $\hat{\beta}_{n,3}$ in (\ref{hatbeta}), $\mathring{\beta}_{n,3}$ in (\ref{ringbeta}), and $\tilde{\beta}_{n,3}$ in (\ref{tildebeta})
for sample size $n=40$ and $n=640$, again for the exponential distribution.
The mean functions take on very small values; the standard deviations are very similar in all cases.

\begin{figure}[htp]
\begin{center}
\includegraphics[scale=0.7]{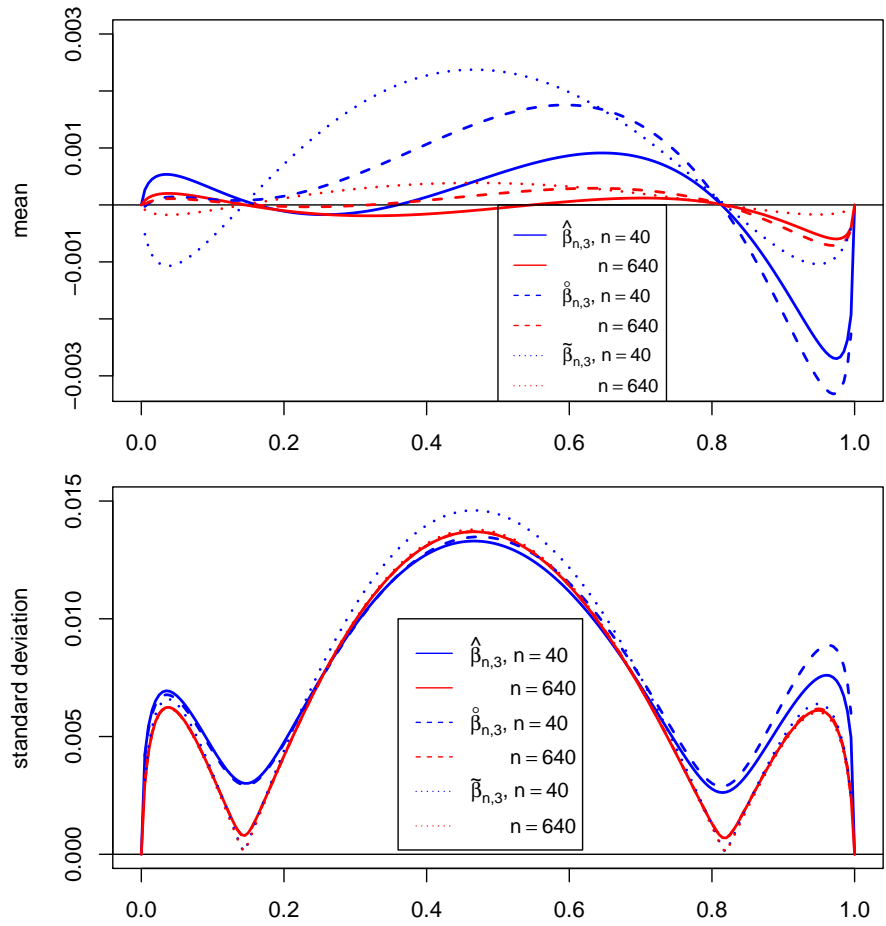}
\caption{\label{figure7} Simulated mean and standard deviation of $\hat{\beta}_{n,3}$, $\mathring{\beta}_{n,3}$
  and $\tilde{\beta}_{n,3}$ for sample size $n=40$ and $n=640$}
\end{center}
\end{figure}

Figure \ref{figure8} shows the function $\tilde{h}_t^{\prime}(\vartheta_0)$, the part inside the brackets
in $\tilde{g}_t^{\prime}(\vartheta_0)$, and $\tilde{g}_t^{\prime}(\vartheta_0)$ itself.
The values of $\tilde{g}_t^{\prime}(\vartheta_0)$ are close to zero on the whole interval. For this reason,
$\hat{\beta}_{n,3}$ is negligible in comparison to $\hat{\beta}_{n,2}$ for the exponential case at hand.

\begin{figure}[htp]
\begin{center}
\includegraphics[scale=0.7]{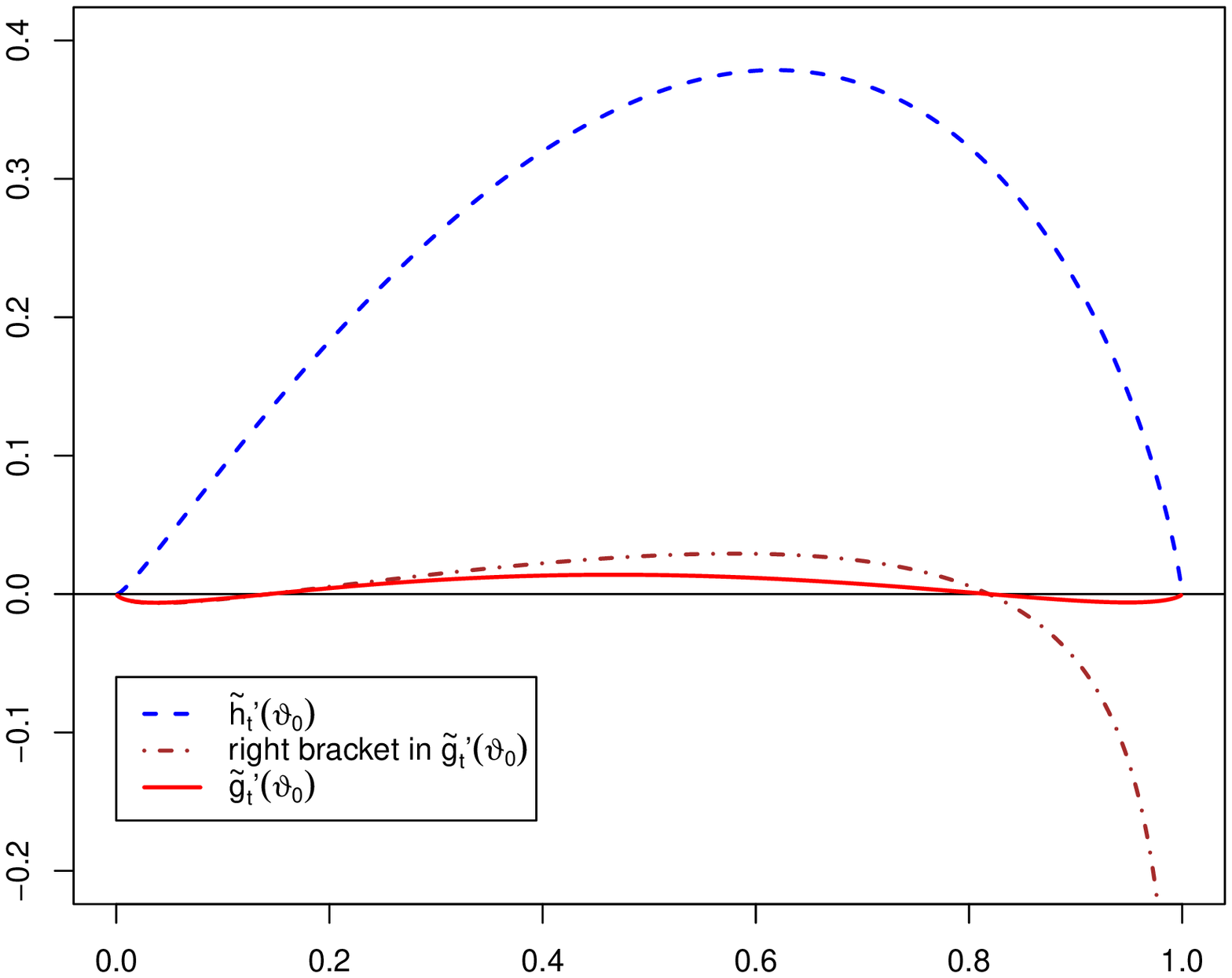}
\caption{\label{figure8} The function $\tilde{h}_t^{\prime}(\vartheta_0)$, the right part of $\tilde{g}_t^{\prime}(\vartheta_0)$,
   and $\tilde{g}_t^{\prime}(\vartheta_0)$ itself}
\end{center}
\end{figure}

We also performed Monte Carlo experiments for other gamma distributions with shape parameter not equal to one. These experiments lead to qualitatively similar results and although not reported here they are available from the authors upon request. A reasonable overall conclusion seems to be that under different sampling scenarios the processes $\hat{\beta}_{n,1}$ and $\hat{\beta}_{n,3}$ in decomposition (A.1) are asymptotically negligible, and hence the behavior of the process $\hat{\beta}_{n}$ of the Chen--Balakrishnan transformation is dominated by the values of the process $\hat{\beta}_{n,2}$. The latter process however coincides with the process $\hat \alpha_n(t)$ of Appendix A.1 which is involved in goodness--of--fit testing for normality with estimated parameters, and this fact justifies the validity of the Chen--Balakrishnan transformation.
\end{appendix}

{}

\end{document}